\documentclass[11pt]{article}
\usepackage[margin=1in]{geometry}
\usepackage{amsmath,amssymb,graphicx,booktabs,hyperref}
\usepackage[numbers,sort&compress]{natbib}
\graphicspath{{figures/}}

\newcommand{\fnone}{\phi_\varnothing}
\newcommand{\coderepo}{\url{https://github.com/hanssmail/darkcorners-abm}}

\title{Herding and Liquidity in Order-Book Markets. I.\\
A Robust Liquidity-Stress Crossover and its Reflexive Mechanism}
\author{Jan Novotny\thanks{The views and opinions expressed in this paper are solely those of the author and do not necessarily reflect those of any affiliated institution.}\\
Centre for Econometric Analysis, Bayes Business School\\
106 Bunhill Row, London EC1Y 8TZ, United Kingdom\\
\texttt{jan@novotny.one}}
\date{\today}

\begin{document}
\maketitle

\begin{abstract}
Agent-based models of markets readily produce emergent instabilities, but telling a
genuine collective effect apart from a parameter artefact takes discipline. We apply
Bouchaud's phase-diagram method to a continuous-double-auction order-book model. The
method is to map the full phase diagram, test its robustness to rule changes, and rule
out degenerate and numerical origins before we call any feature a tipping point. The
model has fundamental-anchored zero-intelligence liquidity and a mid-anchored chartist
herding layer, controlled by the fraction $\varphi$ and the strength $\kappa$ of herders.
A $7\times6$ grid ($336$ runs, each with a scrambled-sign null) locates an emergent
liquidity-stress crossover. The order parameter $\fnone$, the fraction of events with a
one-sided book, rises to $\approx0.34$ at $(\varphi,\kappa)=(0.9,1.0)$, is zero across all
$42$ scrambled cells, and forms a smooth crossover rather than a discontinuous Dark
Corner. The dry-up is rule-robust (it recurs under an order-flow-imbalance rule),
horizon-robust ($\approx0.32$--$0.35$ across a $16\times$ range of momentum window), and
has a monotone onset boundary $\varphi^*(\kappa)=\{0.55,0.45,0.36\}$. We then decompose
the mechanism at a matched directional-bias amplitude
($\mathrm{mean}\,|p_{\rm buy}-0.5|\approx0.269$). Price-momentum herding carries a large,
comparator-robust reflexive component ($+0.29$; buying begets buying), whereas the
order-flow rule's component is $\approx0$ and comparator-dependent. The RMS-mispricing
gradient is a placement artefact, largest at $\kappa=0$. A companion two-market analysis
finds no directional cross-market contagion across a signal-only herding link.
\end{abstract}


\section{Introduction}

Agent-based models (ABMs) of markets and economies are valued because they produce
emergent phenomena that no single agent encodes. That same expressiveness makes them
hard to interpret. A model can be tuned to almost any behaviour, so an apparent
instability may be a genuine collective effect or merely an artefact of a particular
parameter choice. Bouchaud's recent methodological manifesto
\citep{bouchaud2024navigating} sets out a discipline for telling these apart. Rather
than reporting a single calibrated run, one should map the full phase diagram of the
model over its key control parameters, identify the boundaries between qualitatively
distinct regimes, and only then ask whether any boundary is a genuine ``Dark
Corner''--a discontinuity line beyond which runaway instabilities appear. One asks this
by testing the boundary for robustness to rule changes and by ruling out degenerate or
numerical origins. The term ``Dark Corners'' was coined in the macroeconomic Mark-0
lineage \citep{gualdi2016darkcorners,gualdi2015tipping}, and the phase-diagram-first
programme has since been applied to minimal exchange and inequality models
\citep{patil2024inequality}. In all of these the state variable is macroeconomic
(unemployment, wealth) and the mechanism is a firm or household adjustment rule. The same
order-parameter/phase-diagram programme has also been applied to a stylised financial-market
ABM \citep{lye2012orderparams} and, in an order-book setting, to a single-parameter feedback
transition to endogenous liquidity crises \citep{fosset2020liquiditycrises}; in each the
control axes and observables differ from ours and the null/artefact/robustness discipline is
not applied.

To our knowledge, this discipline has not previously been applied to an order-book
microstructure ABM, where the natural control axes are the strength and prevalence of
herding and the breadth of liquidity provision rather than macro policy parameters. The
closest analogue is Gao et al.\ \citep{gao2022flashcrash}, who sweep three microstructure
parameters in a limit-order-book ABM and report a relationship with flash-crash
amplitude. There, however, the crash is triggered by an injected exogenous ``Sell
Algorithm'' calibrated to the historical 6 May 2010 event, and the study neither locates
an endogenous phase boundary nor carries out the artefact-versus-genuine-tipping
discipline that is the actual content of the manifesto. Mapping an order-book ABM's phase
diagram with that discipline is the gap we address.

Mapping the phase diagram and verifying the feature against an explicit null answers
whether the stress is genuine. The manifesto's discipline then asks how robust it is and
what generates it. We therefore test the feature against a change of behavioural rule and
a $16\times$ change of herding horizon, extract its onset boundary, and isolate its
generating feedback with an amplitude-matched open-loop control. Let us isolate that
feedback fairly. The two herding rules must be driven at a matched herd-signal amplitude
before their loops are compared, because their raw signals sit at very different points
of the response nonlinearity. The contrast must also be checked against several open-loop
comparators, since matching only the first moment of the drive leaves its higher-moment
structure unmatched. Carried out this way, price-momentum herding carries a robust,
self-reinforcing reflexive component, positive against every open-loop comparator,
whereas the order-flow-imbalance (OFI) rule's reflexive component is $\approx0$ and
comparator-dependent, with no robust sign.

We map the two-dimensional phase diagram of a validated continuous-double-auction ABM
over its two herding control axes--herd fraction $\varphi$ and herd strength
$\kappa$--apply the full artefact-ruling-out protocol, and then characterise the
resulting feature. We make four contributions, given in order of their strength.
\begin{enumerate}
  \item A genuine, null-verified, imitation-specific liquidity-stress region sits in the
    high-herding corner of the $\varphi\times\kappa$ plane. The order parameter
    $\fnone$--the fraction of events at which the book is one-sided (no two-sided
    mid)--rises to $\approx0.34$ at $(\varphi,\kappa)=(0.9,1.0)$. It is zero across all
    $42$ scrambled-sign null cells, requires both high $\varphi$ and $\kappa>0$, is robust
    to a $4\times$ change of the liquidity-placement bandwidth, and is neither numerical
    nor degenerate. This is a real emergent collective phenomenon.
  \item The existence of the dry-up is robust, which is the headline. It is (a)
    rule-robust: it reappears under a completely different herding rule, order-flow
    imbalance (OFI), reaching $\fnone=0.227$ in the same corner with the scrambled null
    still identically zero across all cells. It is (b) horizon-robust:
    $\fnone\approx0.32$--$0.35$ across momentum windows $w\in\{50,\dots,800\}$, fading
    only when the momentum response is deliberately weakened, exactly as a genuine
    destabilising mechanism should. And it is (c) endowed with a clean, monotone onset
    boundary $\varphi^*(\kappa)=\{0.55,0.45,0.36\}$ at $\kappa=\{0.6,0.8,1.0\}$: stronger
    herding lowers the onset fraction, with a smooth (no-jump) rise whose monotone shape
    is invariant across $\fnone$ thresholds spanning $0.005$--$0.05$.
  \item Price-momentum herding carries a self-reinforcing reflexive component. We isolate
    each rule's reflexive feedback with an amplitude-matched open-loop control. The two
    rules are driven at the same closed-loop directional-bias amplitude
    ($\mathrm{mean}\,|p_{\rm buy}-0.5|\approx0.269$), and the reflexive
    component--closed-loop ``real'' minus an amplitude-matched open-loop drive--is
    measured for each against several comparators. Price momentum carries a large,
    comparator-robust, self-reinforcing component ($+0.29$): positive against the
    open-loop shadow ($+0.29$), against the loop-isolating replay ($+0.21$), and far above
    every synthetic-telegraph cell. The closed loop thus manufactures $\sim7\times$ more
    dry-up than any equally-forceful exogenous drive: the herd reads a mid it is itself
    moving, and buying begets buying. Under OFI, by contrast, the reflexive component is
    $\approx0$ and comparator-dependent, with no robust sign. It is $-0.22$ only against
    one baseline-flow shadow (a $\sim2\times$ outlier among matched-amplitude drives), but
    $+0.03$ against the loop-isolating replay and $\approx0$ against a persistence-matched
    telegraph and the continuum grid. Matching the herd-side first-moment amplitude does
    not match the drive's higher-moment (burst) structure, which is why a fair measurement
    needs several comparators, and why the two rules' components are not symmetric
    opposite-signed twins (Section~\ref{sec:mechanism}). The dry-up's magnitude,
    separately, follows a smooth amplitude$\times$persistence continuum.
  \item The feature is a smooth crossover, not a sharp Dark Corner, confirmed at fine
    resolution ($\Delta\varphi=0.05$ over three $\kappa$ rows). $\fnone$ climbs
    continuously and saturates at the grid corner; we find no discontinuity line, runaway,
    or crash, and do not claim a tipping line. The RMS-mispricing growth toward high
    $\varphi$ is a placement/dilution artefact--largest at $\kappa=0$ and suppressed by
    real imitation--which we decline to call a Dark Corner.
\end{enumerate}

In sum, the model shows an emergent collective effect that survives an explicit null, a
rule- and horizon-robustness battery, and an amplitude-matched open-loop mechanism
decomposition. It is a robust liquidity-stress crossover with a signed, rule-dependent
reflexive mechanism, rather than a universal tipping point. As a corollary, the
self-referential mechanism predicts that the stress is intrinsically local: a companion
two-market analysis finds no directional cross-market contagion through signal-only
herding, since an external open-loop signal cannot instantiate a market's own reflexive
loop (Section~\ref{sec:contagion}).

\section{Related Work}

Our method is taken directly from Bouchaud \citep{bouchaud2024navigating}. One maps the
ABM's phase diagram over its control parameters first, treats ``Dark Corners'' as
discontinuity lines beyond which runaway instabilities appear, and tests the robustness
of any such feature before believing it. The approach descends from the Mark-0
macroeconomic lineage, where a genuine good-economy/bad-economy tipping point was located
and its robustness checked against model variants \citep{gualdi2015tipping}, and where the
term ``Dark Corners'' was coined for policy-triggered instability boundaries
\citep{gualdi2016darkcorners}. More recently the same phase-transition discipline was
applied to a minimal good-exchange model in which market clearing becomes dynamically
unreachable beyond a threshold \citep{patil2024inequality}. In every case the state
variable is macroeconomic. We import the identical discipline into an order-book
microstructure setting.

Herding strength as a control parameter that drives endogenous critical or intermittent
dynamics is a classic theme. The Lux--Marchesi chartist/fundamentalist model
\citep{lux1999scaling} and the Cont--Bouchaud herding mechanism \citep{cont2000herd} are
the direct ancestors of the herding ingredient in our vehicle, and both motivate herd
fraction and strength as the natural control axes. The Minority-Game literature provides
the closest classical precedent for treating a market-game control parameter as tuning a
genuine collective phase transition with an identifiable order parameter
\citep{marsili2001continuum}. Filimonov and Sornette \citep{filimonov2012reflexivity}
frame instability as distance to an endogeneity/criticality threshold. This sharpens the
distinction, central to our work, between a genuine endogenous collective effect and an
exogenously injected shock.

The nearest prior work is Gao et al.\ \citep{gao2022flashcrash}, a limit-order-book
microstructure ABM that sweeps three control parameters (sell-algorithm volume share,
market-maker inventory limit, fundamental-trader frequency) and reports functional
relationships with simulated flash-crash amplitude. A related endogenous order-book flash
model is Kårvik et al.\ \citep{karvik2018deeds}, whose sterling--dollar ABM produces flash
episodes from the procyclical withdrawal of high-frequency participants, but as a
single-parameter sensitivity study (episode frequency versus fast-trader share) rather than
a phase-diagram or onset-boundary map. This is the closest prior work to ``a
phase diagram of an order-book ABM.'' We differ on three explicit points. First, on
mechanism: their crash is driven by an injected exogenous shock, a Sell Algorithm
mimicking the historical 6 May 2010 flash crash, whereas ours is endogenous, since the
liquidity stress emerges purely from the interaction of directional imitation with a
finite order book and with no exogenous trigger. Second, on the object of study: they
calibrate to reproduce one historical crash's amplitude, whereas we locate a region
boundary in a two-dimensional intrinsic parameter plane, as the phase-diagram-first
prescription requires. Third, on integrity discipline: they do not separate genuine
emergent tipping from degenerate or numerical-artefact instability, nor do they test null
and robustness controls. Our central methodological addition is exactly this discipline: a
scrambled-sign null, a liquidity-bandwidth robustness cut, and numerical-stability checks.
The outcome is that the effect is a smooth crossover, not a discontinuous crash.

The closest econophysics precedents map phases without the microstructure-plus-discipline
combination we use. Lye, Tan and Cheong \citep{lye2012orderparams} construct a
two-dimensional phase diagram of a stylised multi-asset market ABM from steady-state order
parameters (dead/boom/jammed phases), but their price mechanism is a clearing-house
adjustment rule rather than a bid--ask limit order book, their control axes are random
buy/sell fractions rather than herding and liquidity, and they apply no null or
artefact-versus-genuine-tipping discipline. Fosset, Bouchaud and Benzaquen
\citep{fosset2020liquiditycrises} exhibit a genuine second-order phase \emph{transition} to
endogenous liquidity crises in a stylised order book driven by a single feedback-intensity
parameter; ours instead maps a two-dimensional herding$\times$liquidity plane, tests it
against an explicit null and robustness battery, and finds a smooth crossover rather than a
critical point. We are not aware of prior work combining a genuine order-book microstructure
ABM, a two-dimensional herding/liquidity phase diagram, and the full artefact-ruling-out
discipline in this sense. Finally, sloppy/stiff parameter-direction analysis
\citep{waterfall2006sloppy}, which the manifesto imports for high-dimensional navigation,
enters here only qualitatively (Section~\ref{sec:onset}: the stiff directions are the two
herding axes, the horizon directions are sloppy); a formal sensitivity-spectrum treatment
is left to future work once more than two axes are in play.

\section{Model}

\subsection{Market model}
The market is a continuous double-auction (CDA) limit-order book populated by
fundamental-anchored zero-intelligence (ZI) liquidity traders and a chartist
\emph{herding} layer designed to inject positive feedback. Herders read a
price-momentum signal computed over a rolling window and place \emph{mid-anchored}
orders--i.e.\ they chase the current mid price, in the destabilising
DeLong/Lux--Marchesi sense \citep{lux1999scaling,cont2000herd}. The model is a fair
vehicle for a phase-diagram study on three counts: (i) its order book passes the
standard microstructure validation gates (the ZI baseline reproduces the expected
spread, depth, impact and diffusion behaviour); (ii) its herding layer exposes exactly
the two control knobs used as phase-diagram axes; and (iii) the chartist's imitation
channel is genuine rather than built in--an ablation ladder with a scrambled-null
decomposition confirms a directional imitation channel, ruling out the ``result built
into the model'' failure mode.

\subsection{Control parameters}
The two axes of the phase diagram are:
\begin{itemize}
  \item $\varphi \in [0,1]$: the \emph{fraction} of agents that are herders (vs.\ ZI
    liquidity providers);
  \item $\kappa \ge 0$: the \emph{strength} of the herding response to the momentum
    signal.
\end{itemize}
Setting $\kappa=0$ turns off directional herding: a nominal ``herder'' then places a
random-side, mid-anchored order, so the $\kappa=0$ column isolates the pure
composition/placement effect of replacing fundamental-anchored ZI orders with
mid-anchored random-side ones, with \emph{no} directional imitation.

\subsection{Two herding rules}
\label{sec:rules}
The robustness of the phenomenon to a change of behavioural rule is central to the
phase-diagram discipline, so the herd layer is run under \emph{two} alternative
directional signals:
\begin{itemize}
  \item Price momentum (the default): the herd side follows the sign of the price change
    over a rolling window $w$; herders chase the current mid in the destabilising
    DeLong/Lux--Marchesi sense \citep{lux1999scaling,cont2000herd}.
  \item Order-flow imbalance (OFI): the herd side follows the sign of the realised
    buy/sell trade-flow imbalance over the window. This is a flow-chasing rule that reads
    the trade tape rather than the mid.
\end{itemize}
Momentum reads a price the herd itself moves; OFI reads realised order flow. This
distinction becomes the key to the mechanism analysis (Section~\ref{sec:mechanism}):
once the two rules are compared at a matched herd-signal amplitude, only momentum's loop
carries a robust self-reinforcing reflexive component, while OFI's is
comparator-dependent and has no robust sign.

\subsection{The scrambled-sign null}
Our central control is a \emph{scrambled-sign null}. In the null condition the herd
layer is run with the \emph{same} order-type mix and the \emph{same} $\varphi$-driven
reduction of ZI arrivals, but the \emph{sign} (buy/sell direction) of each herd order
is randomised, destroying directional correlation while preserving everything else.
Any phenomenon that survives in the real condition but vanishes in the null is
therefore attributable to \emph{directionally correlated imitation}, not to the
mechanical dilution of ZI order flow. Because herders are reactive, we use a unique,
never-reused seed for every (real/null) condition and never share common random
numbers across conditions--common random numbers would spuriously couple the
conditions through the reactive herd response.

\subsection{Observables}
All order parameters are computed on a \emph{continuous} reference-price series. The
key observables are:
\begin{itemize}
  \item $\fnone$: the fraction of events at which the book has no
    two-sided mid (one side has emptied)--an emergent liquidity dry-up indicator,
    and the order parameter that separates the stressed from the
    quiescent regime;
  \item $\mathrm{RMS}|{\rm mid}-{\rm fund}|$: root-mean-square deviation of the
    reference price from the random-walking fundamental value;
  \item return volatility (mid-to-mid, sampling interval $\mathrm{d}t=25$ events),
    peak $|{\rm mid}-{\rm fund}|$ excursion, max drawdown per $1000$-event window, and
    the signed mean deviation (to test for bubble/crash directionality).
\end{itemize}

Let us make the order-parameter definition consistent across conditions. A naive ``drop
events where the mid is undefined'' rule discards precisely the most-stressed events from
the real condition (where $\fnone\to0.34$) while discarding none from the null (where
$\fnone=0$), which biases any cross-condition comparison. We therefore fall back to the
last traded price whenever the mid is undefined and carry it forward. The traded price
stays defined during one-sided stretches, since aggressive orders keep executing. This
gives a consistent order-parameter definition across the compared conditions.

\subsection{Baseline validation}
At $(\varphi,\kappa)=(0,0)$ over $150{,}000$ events, the mid tracks the
random-walking fundamental, the mean spread is $\approx3.1$ ticks (stationary),
$\fnone=0$ (always two-sided), and $\mathrm{RMS}|{\rm mid}-{\rm
fund}|\approx2.6$--consistent with the validated ZI foundation. The foundation
gate is passed: both a stable/efficient regime (e.g.\ $(\varphi,\kappa)=(0.1,0.2)$:
spread $\approx3.1$, $\fnone=0$) and a genuinely stressed regime (e.g.\
$(0.9,1.0)$: spread $\approx1.75$, $\fnone\approx0.34$) are reachable in
parameter space, so the model is a fair vehicle in which to search for a phase
boundary.

\section{Simulation methods}

\subsection{Phase-diagram sweep}
The main experiment sweeps the grid
\[
  \varphi \in \{0,\,0.15,\,0.3,\,0.45,\,0.6,\,0.75,\,0.9\}
  \quad\times\quad
  \kappa \in \{0,\,0.2,\,0.4,\,0.6,\,0.8,\,1.0\},
\]
i.e.\ $7\times6 = 42$ cells. Each cell is run with $5$ real ($\mathrm{scrambled}=\mathrm{False}$)
and $3$ scrambled-sign null ($\mathrm{scrambled}=\mathrm{True}$) independent seeds, for
$42\times8 = 336$ runs in total. A unique, never-reused seed is assigned per
$(\varphi,\kappa,\text{condition},\text{seed})$; no common random numbers are shared
across conditions.

Each run uses the continuous double-auction book with its validated ZI-baseline
parameters (market-order probability $p_{\text{market}}=0.15$, cancellation
probability $p_{\text{cancel}}=0.05$, limit-order placement bandwidth $b=4$, initial
reference price $\mathrm{ref}_0=1000$, fundamental volatility $\sigma_f=0.05$) and a
herding layer with momentum window $w=200$ and response scale $2.0$, for
$n_{\text{events}}=150{,}000$ events with a burn-in of $20{,}000$ events.

\subsection{Robustness and mechanism experiments}
\label{sec:iter2exp}
On top of the base phase diagram we run a battery of experiments that test the
robustness of the liquidity-stress feature and isolate its mechanism. All hold the
base defaults ($n_{\text{events}}=150{,}000$, burn-in $20{,}000$, the ZI-baseline
parameters above, $w=200$, response scale $2.0$ unless swept); every condition carries
its own never-reused seeds and a scrambled-sign null.
\begin{itemize}
  \item The rule-robustness grid re-runs the herd layer with the OFI signal
    (Section~\ref{sec:rules}) over $\varphi\in\{0.3,0.45,0.6,0.75,0.9\}\times
    \kappa\in\{0.4,0.6,0.8,1.0\}$, $3$ real $+\,2$ scrambled seeds per cell ($100$
    runs). The $\varphi=0$ and $\kappa=0$ rows are dropped as uninformative for a rule
    test (no herder acts / no directional signal).
  \item The horizon-robustness sweep runs at the corner $(0.9,1.0)$ and an onset point
    $(0.6,0.8)$. We sweep the momentum window $w\in\{50,100,400,800\}$ (scale fixed) and
    the tanh $\text{scale}\in\{0.5,1,4,8\}$ ($w$ fixed), $3$ real $+\,1$ scrambled seeds
    per cell ($64$ runs); the $(200,2.0)$ baseline is taken from the main sweep.
  \item The onset-boundary scan is a fine $\varphi$ scan ($\Delta\varphi=0.05$)
    bracketing the transition at $\kappa\in\{0.6,0.8,1.0\}$, $3$ real $+\,1$ scrambled
    seeds per cell ($84$ runs). It resolves the onset boundary $\varphi^*(\kappa)$ and
    tests smoothness at $3\times$ the main grid's $\varphi$ resolution.
  \item The amplitude-matched shadow / replay decomposition isolates the reflexive
    feedback each rule generates, controlling for the herd-signal-amplitude confound
    described in Section~\ref{sec:mechanism}. At the corner we compare three conditions
    for each rule: \textsc{real} (closed loop, in which the herd side follows the market's
    own signal); \textsc{shadow} (open loop, in which the herd side is driven by the
    signal of an independent $\varphi=0$ market, so placement is still anchored to the
    fresh market's mid but the signal is severed from its price); and \textsc{replay} (a
    closed-loop-generated signal replayed open-loop into a fresh market). Both the
    closed-loop leg and the shadow driving series are calibrated to a matched herd-side
    directional-bias amplitude, $\mathrm{mean}\,|p_{\rm buy}-0.5|\approx0.269$. Momentum
    is kept at its published scale $2.0$ (the reference target), and OFI's tanh scale is
    bisected directly in the closed loop to the same amplitude (converged scale $68.8$).
    The reflexive component is then $\textsc{real}-\textsc{shadow}$ at matched amplitude,
    reported per rule with SEM. We use $\ge6$ seeds per condition and $\ge3$ independent
    driving series for the shadow leg ($3$ independent $\varphi=0$ baseline markets
    $\times2$ seeds), and one consistent replay burn-in convention for both rules.
  \item The amplitude$\times$persistence continuum tests whether the dry-up's magnitude
    is a smooth function of an exogenous drive's statistics. A synthetic two-state
    (telegraph) exogenous drive replaces the herd side at the corner, with amplitude
    $\in\{0.13,0.269,0.40\}$ and mean run-length (persistence) $\in\{8,30,60,110\}$ events
    set independently of each other ($48$ runs, $4$ seeds/cell). The persistence range
    spans momentum's natural raw-signal persistence ($\approx7.4$--$7.8$ events) up to
    OFI's ($\approx99$--$134$ events).
\end{itemize}

\subsection{Artefact-ruling-out controls}
Beyond the scrambled-sign null (built into every cell), we run two additional controls.
The first is a liquidity-bandwidth robustness cut. At $\kappa=1.0$, we re-run
$\varphi\in\{0.45,0.6,0.75,0.9\}$ while varying the ZI limit-order placement bandwidth
$b\in\{2,4,8\}$. The bandwidth sets the scatter width of resting limit orders, i.e.\ the
breadth of provided liquidity. If $\fnone$ were an artefact of one specific liquidity
setting, a $4\times$ change of $b$ would move it. The onset row ($\varphi=0.45$) is
included to probe band-sensitivity near onset; the RMS metric in this cut uses the same
continuous-reference definition as the main sweep (Section~\ref{sec:artefact}). The second
control is a numerical-stability audit. We check every run (the $336$ base runs and the
$\sim320$ robustness runs) for non-finite values, proximity of prices to the hard floor,
and whether the fundamental anchor remains active throughout.

\section{Results}

\subsection{A null-verified, imitation-specific liquidity-stress region}

The order parameter that separates the stressed from the quiescent regime is
$\fnone$, the fraction of events at which one side of the book has
emptied. Table~\ref{tab:fracnone} gives its mean over the $5$ real seeds (SEM
$\le0.008$ everywhere).

\begin{table}[htbp]
\centering
\caption{$\fnone$ (fraction of events with a one-sided book), REAL
condition, mean over $5$ seeds. Rows: herd strength $\kappa$; columns: herd fraction
$\varphi$. The stress region is the high-$\varphi$/high-$\kappa$ corner; it requires
\emph{both} axes. In the scrambled-sign null, every one of these $42$ cells is
$0.000$.}
\label{tab:fracnone}
\begin{tabular}{c|ccccccc}
\toprule
$\kappa \backslash\, \varphi$ & 0.0 & 0.15 & 0.3 & 0.45 & 0.6 & 0.75 & 0.9 \\
\midrule
1.0 & 0 & 0 & 0 & 0.083 & 0.193 & 0.275 & \textbf{0.340} \\
0.8 & 0 & 0 & 0 & 0.022 & 0.128 & 0.222 & 0.288 \\
0.6 & 0 & 0 & 0 & 0.001 & 0.051 & 0.147 & 0.227 \\
0.4 & 0 & 0 & 0 & 0     & 0     & 0.040 & 0.129 \\
0.2 & 0 & 0 & 0 & 0     & 0     & 0     & 0.001 \\
0.0 & 0 & 0 & 0 & 0     & 0     & 0     & 0     \\
\bottomrule
\end{tabular}
\end{table}

$\fnone$ rises to $\approx0.34$ at $(\varphi,\kappa)=(0.9,1.0)$--the
book is one-sided about a third of the time--and forms a diagonal region that
requires \emph{both} high $\varphi$ and $\kappa>0$: it is $\approx0$ for $\kappa\le0.2$
at any $\varphi$ and stays below $0.2\%$ for $\varphi\le0.3$ at any $\kappa$ (it is
\emph{not} identically zero there--e.g.\ $\fnone=0.002$ at
$(0.30,1.0)$ in the fine onset scan of Section~\ref{sec:onset}; the tabulated $0$s are
roundings, and this small nonzero tail \emph{supports} the smooth-crossover
reading, with no hard threshold). The onset $\varphi$ moves down as $\kappa$ rises
(from $\varphi\approx0.75$ at $\kappa=0.4$ to $\varphi\approx0.45$ at
$\kappa=0.8$--$1.0$; sharpened to a boundary $\varphi^*(\kappa)$ in
Section~\ref{sec:onset}). This is the clean positive result: an emergent liquidity
stress that appears only when a large fraction of strongly imitating agents interacts
with the finite book.

\begin{figure}[htbp]
\centering
\includegraphics[width=\textwidth]{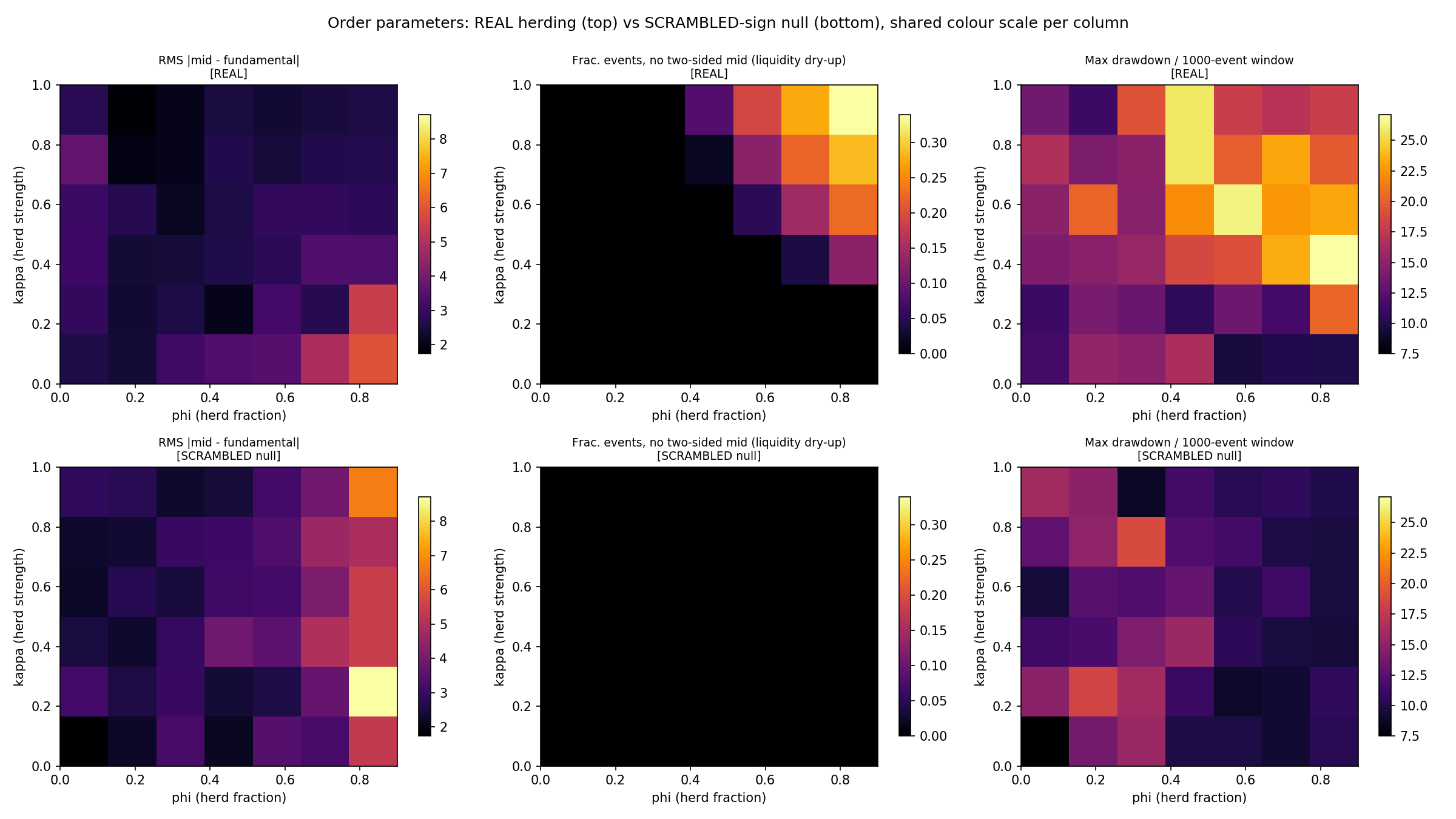}
\caption{Order parameters, REAL herding (top) vs.\ SCRAMBLED-sign null (bottom),
shared colour scale per column. \emph{Middle column} ($\fnone$): the
liquidity dry-up is a bright ramp in the high-$\varphi$/high-$\kappa$ corner for the
real condition and is \emph{identically zero} everywhere in the null--the key null
contrast. \emph{Left column} (RMS mispricing): the gradient is present in both
conditions and, if anything, stronger in the null (see
Section~\ref{sec:artefact}). \emph{Right column} (max drawdown): noisy, with no clean
structure and real $\approx$ null. Axes: $\varphi$ (herd fraction) horizontal,
$\kappa$ (herd strength) vertical.}
\label{fig:realscrambled}
\end{figure}

The decisive control is the scrambled-sign null. Here $\fnone=0$ across the entire null
plane (all $42$ cells, max $=0.000$). The null keeps the same order-type mix and the same
$\varphi$-driven reduction of ZI arrivals, but randomises the herd side. The dry-up
appears only when herd orders are directionally correlated, so it is a genuine
collective/imitation effect, not a consequence of ``fewer ZI orders.''
Figure~\ref{fig:realscrambled} shows this: the middle column has a bright ramp in the
top-right for the real condition and an entirely black panel for the null.

The saturated corner is band-invariant, while the onset is mildly band-sensitive. At
$\kappa=1.0$, varying the liquidity-placement bandwidth $b\in\{2,4,8\}$ barely moves
$\fnone$ in the developed region ($\varphi\ge0.6$; Table~\ref{tab:band}). A $4\times$
change of the breadth of provided liquidity leaves the saturated dry-up essentially
unchanged, so it is emergent from the herding dynamics rather than an artefact of one
liquidity setting. The onset, however, is mildly bandwidth-sensitive: at $\varphi=0.45$,
$\fnone$ falls from $0.131$ (at $b=2$) to $0.067$ (at $b=8$), a $\sim2\times$ swing, as a
wider placement scatter delays onset. This fits the picture that the onset boundary is the
delicate part of the phase diagram (Section~\ref{sec:onset}) while the corner is not. We
therefore state invariance for the saturated corner and band-sensitivity near onset,
rather than a blanket ``band-invariant''.

\begin{table}[htbp]
\centering
\caption{Band-sensitivity of $\fnone$ at $\kappa=1.0$
(mean$\pm$SEM). A $4\times$ change of the placement bandwidth barely moves the order
parameter in the saturated region ($\varphi\ge0.6$) but shifts it by $\sim2\times$ at
the onset ($\varphi=0.45$).}
\label{tab:band}
\begin{tabular}{c|ccc|l}
\toprule
$\varphi$ & $b=2$ & $b=4$ & $b=8$ & reading \\
\midrule
$0.45$ (onset)  & $0.131\pm0.005$ & $0.084\pm0.004$ & $0.067\pm0.006$ & bandwidth-\emph{sensitive} ($\sim2\times$) \\
0.6  & $0.221\pm0.004$ & $0.195\pm0.002$ & $0.203\pm0.002$ & invariant \\
0.75 & $0.288\pm0.003$ & $0.278\pm0.004$ & $0.285\pm0.003$ & invariant \\
0.9 (corner) & $0.343\pm0.005$ & $0.337\pm0.002$ & $0.348\pm0.003$ & invariant \\
\bottomrule
\end{tabular}
\end{table}

The stress is neither numerical nor degenerate. No non-finite value occurs in any of the
$336$ base runs or the $\sim320$ robustness runs (max $\mathrm{RMS}=17.0$, finite); prices
never approach the hard floor of $1$ (observed minimum $\approx942$); and the fundamental
anchor ($\sigma_f=0.05$) is active throughout. The stress is a book-sidedness phenomenon
at a price that stays near fundamental, not a blowup. It is also symmetric: the signed
mean deviation at every corner cell is $\approx0$ (its absolute value below $0.2$ ticks;
final deviation within $\pm6$). The stressed market does not run away up (bubble) or down
(crash). The price stays pinned near the fundamental while the book repeatedly goes
one-sided. This is a liquidity-stress phenomenon, not a directional price tipping point.

\subsection{A smooth crossover, not a sharp Dark Corner}

\begin{figure}[htbp]
\centering
\includegraphics[width=\textwidth]{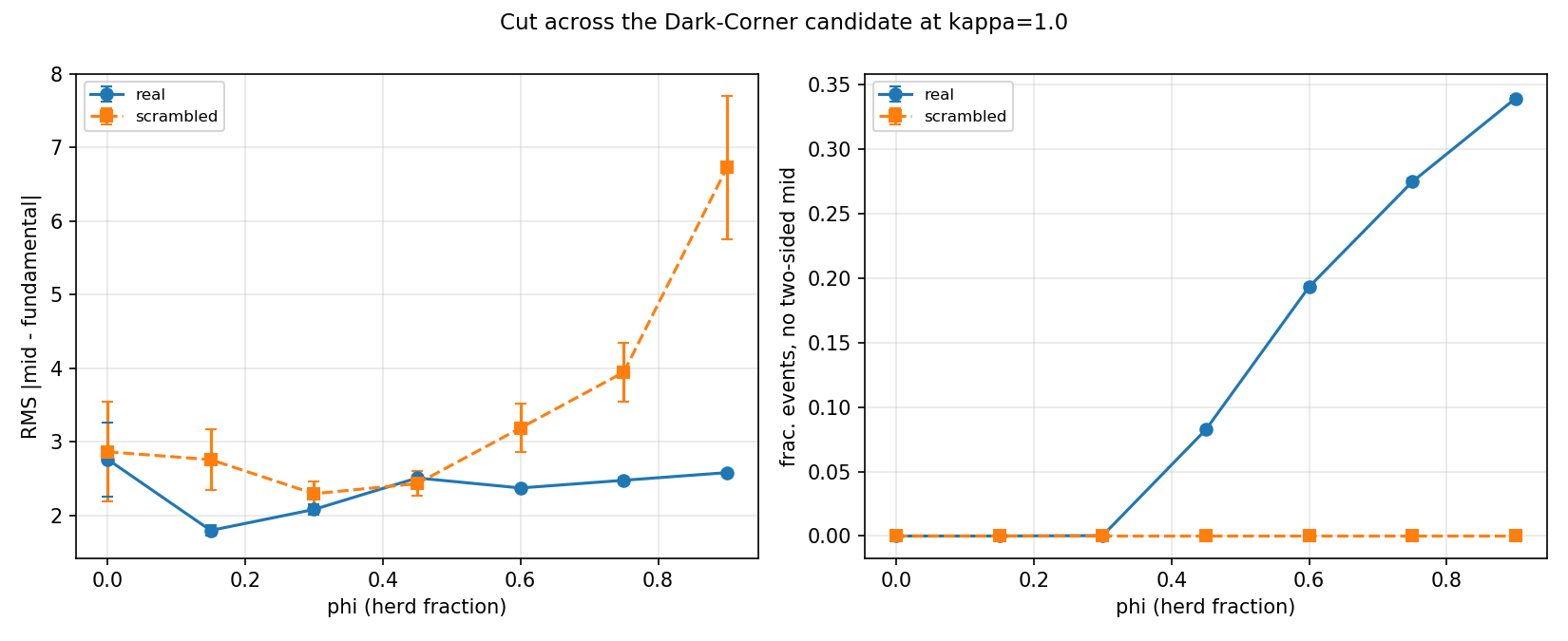}
\caption{Cut across the high-herding corner at $\kappa=1.0$, real vs.\ scrambled null
(mean$\pm$SEM). \emph{Right:} $\fnone$ rises smoothly and monotonically
for the real condition (blue) from $\varphi\approx0.3$ onward, saturating at
$\approx0.34$; the null (orange) is flat at zero--a smooth, imitation-specific
crossover. \emph{Left:} RMS mispricing--the null (orange) rises to $\approx6.7$
at $\varphi=0.9$ while the real curve (blue) stays flat at $\approx2.5$, showing that
directional imitation \emph{suppresses}, not amplifies, mispricing.}
\label{fig:cut}
\end{figure}

Along the $\kappa=1.0$ cut, $\fnone$ climbs continuously and monotonically--$0 \to 0.083
\to 0.193 \to 0.275 \to 0.340$--and saturates near the grid corner (Fig.~\ref{fig:cut},
right panel). There is no sign of a discontinuous transition.
\begin{itemize}
  \item There is no discontinuity line: cell-to-cell increments are smooth
    ($\approx0.05$--$0.10$ per grid step, each well outside SEM but with no jump).
  \item There is no runaway and no crash: the price stays near fundamental and max
    drawdown does not diverge.
  \item There is no order-parameter jump of the kind a Dark Corner denotes.
\end{itemize}

In this vehicle and grid there is no genuine discontinuous Dark Corner or tipping line.
What exists is a smooth, imitation-driven liquidity-stress crossover in the
high-$\varphi$/high-$\kappa$ corner, which we report as a crossover rather than a tipping
point. Finer $\varphi$ scans that exclude a sub-cell discontinuity (Section~\ref{sec:onset},
$\Delta\varphi=0.05$ over three $\kappa$ rows, with an independent $\Delta\varphi=0.03$
scan in agreement) confirm the smooth ramp at finer resolution.

\subsection{Rule-robustness: the dry-up survives the order-flow-imbalance signal}
\label{sec:rule}
\begin{figure}[htbp]
\centering
\includegraphics[width=\textwidth]{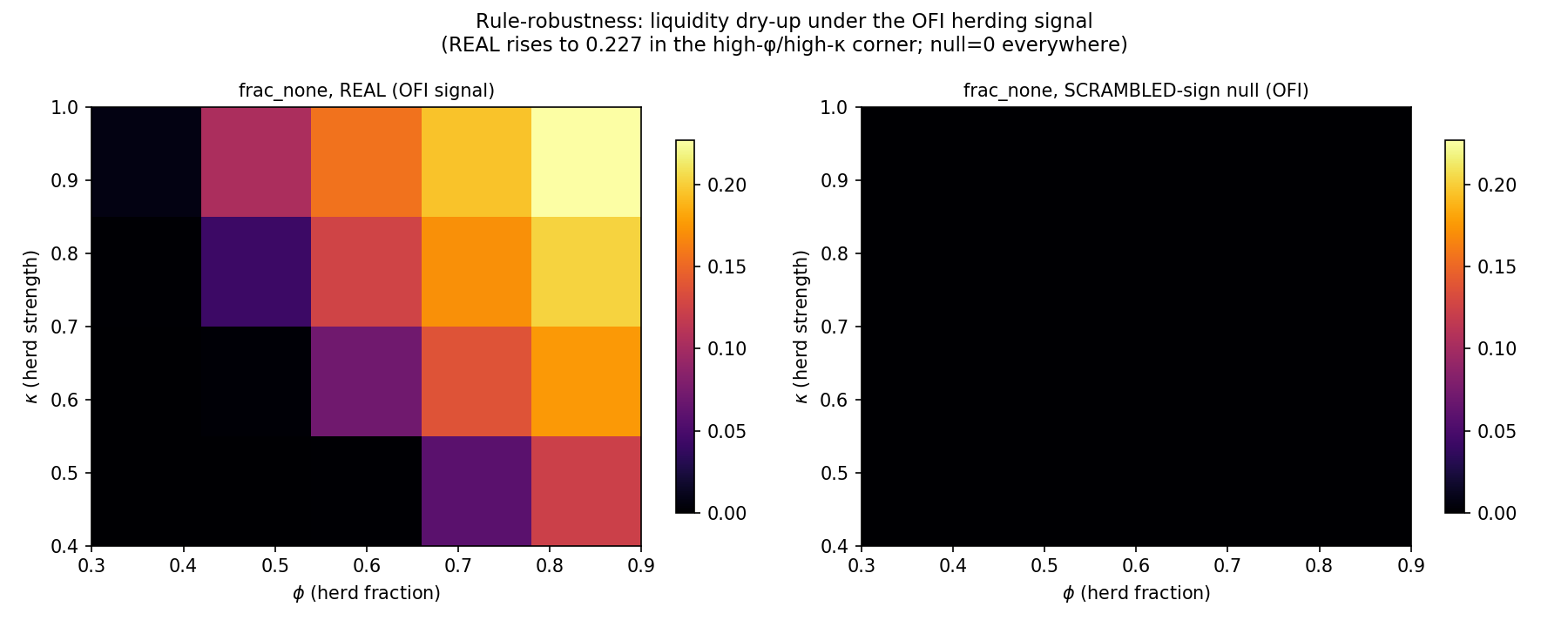}
\caption{Rule-robustness under the OFI herding signal. \emph{Left:} REAL
$\fnone$ rises to $0.227$ in the high-$\varphi$/high-$\kappa$ corner, the
same diagonal region found under price momentum. \emph{Right:} the scrambled-sign null
is identically zero across the whole plane--the dry-up requires directionally
correlated OFI-driven herding, exactly as under momentum. Axes as in
Fig.~\ref{fig:realscrambled}.}
\label{fig:ofi}
\end{figure}

The manifesto's first robustness test is a change of behavioural rule. We replace the
price-momentum signal with order-flow imbalance (OFI; Section~\ref{sec:rules}), a
different heuristic that reads the trade tape rather than the mid. The same phenomenon
appears (Table~\ref{tab:ofi}, Fig.~\ref{fig:ofi}). $\fnone$ rises to $0.227$ in the same
high-$\varphi$/high-$\kappa$ corner and occupies the same diagonal region, requiring both
high $\varphi$ and $\kappa$, with the onset $\varphi$ falling as $\kappa$ rises.
Decisively, the scrambled-sign null is identically zero across all $20$ cells. The corner
dry-up is weaker than momentum's $0.34$, but not because OFI is a gentler per-order rule.
The opposite is true: at the shared scale the OFI signal saturates the tanh, so per order
the OFI herd is more aggressive (near-deterministic side-following, $\mathrm{mean}\,|p_{\rm
buy}-0.5|\approx0.47$ vs momentum's $\approx0.26$; Section~\ref{sec:mechanism}). The weaker
corner is a property of the closed-loop dynamics, not of rule gentleness: the mechanism
decomposition in Section~\ref{sec:mechanism} finds the OFI loop's reflexive component to be
$\approx0$ and comparator-dependent, with no robust sign, rather than self-reinforcing like
momentum's. What matters here is that the phenomenon and its null both transfer, so the
liquidity-stress crossover is not an artefact of the price-momentum heuristic.

\begin{table}[htbp]
\centering
\caption{$\fnone$ under the OFI herding rule, REAL condition, mean over
$3$ seeds (SEM $\le0.008$). The scrambled-sign null is $0.000$ across all $20$ cells.
The same corner region as momentum appears, at weaker magnitude.}
\label{tab:ofi}
\begin{tabular}{c|ccccc}
\toprule
$\kappa \backslash\, \varphi$ & 0.3 & 0.45 & 0.6 & 0.75 & 0.9 \\
\midrule
1.0 & 0.007 & 0.105 & 0.156 & 0.195 & \textbf{0.227} \\
0.8 & 0.002 & 0.043 & 0.127 & 0.172 & 0.203 \\
0.6 & 0.000 & 0.002 & 0.071 & 0.138 & 0.176 \\
0.4 & 0.000 & 0.000 & 0.002 & 0.059 & 0.123 \\
\bottomrule
\end{tabular}
\end{table}

\subsection{Horizon-robustness}
\label{sec:horizon}
\begin{figure}[htbp]
\centering
\includegraphics[width=\textwidth]{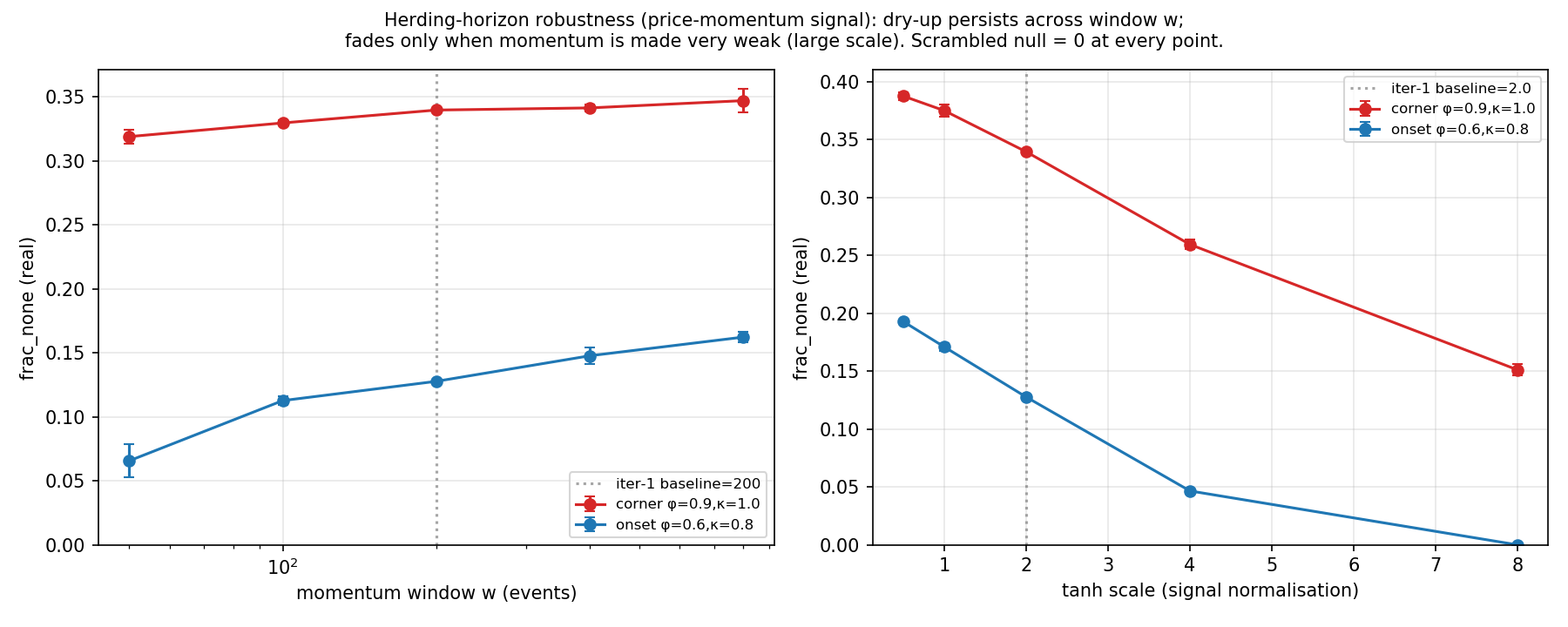}
\caption{Horizon-robustness (price-momentum signal), corner $(0.9,1.0)$ and onset
$(0.6,0.8)$; scrambled null $=0$ at every point. \emph{Left:} across a $16\times$ range
of momentum window $w$ the corner dry-up is flat at $0.32$--$0.35$, i.e.\ horizon-robust.
\emph{Right:} across the tanh scale, the dry-up strengthens when the signal is
sharpened (small scale) and \emph{fades} when it is weakened (large scale $\to0$ at the
onset), as a genuine destabilising mechanism should.}
\label{fig:horizon}
\end{figure}

The effect is also not a peculiarity of one herding horizon (Fig.~\ref{fig:horizon}).
Sweeping the momentum window $w$ over a $16\times$ range at the corner leaves
$\fnone$ essentially flat--$0.319,0.330,0.340,0.341,0.347$ for
$w=50,100,200,400,800$ (scrambled $=0$ at every $w$)--with a mild increase at the
onset point $(0.6,0.8)$. It \emph{is} scale-dependent in the expected direction: the
tanh $\text{scale}$ sets how strongly the herd reacts (smaller scale $=$ sharper
response $=$ stronger positive feedback), and the dry-up strengthens as the signal is
made stronger (corner $0.388$ at scale $0.5$) and \emph{fades as it is weakened}
(corner $0.151$ at scale $8$; the onset point falls to $\approx0$). Turning the
destabilising response down turns the phenomenon off--the correct sign for a genuine
destabilisation-driven effect, not a numerical artefact of one horizon.

\subsection{The onset boundary \texorpdfstring{$\varphi^*(\kappa)$}{phi*(kappa)}: smooth and monotone}
\label{sec:onset}
\begin{figure}[htbp]
\centering
\includegraphics[width=\textwidth]{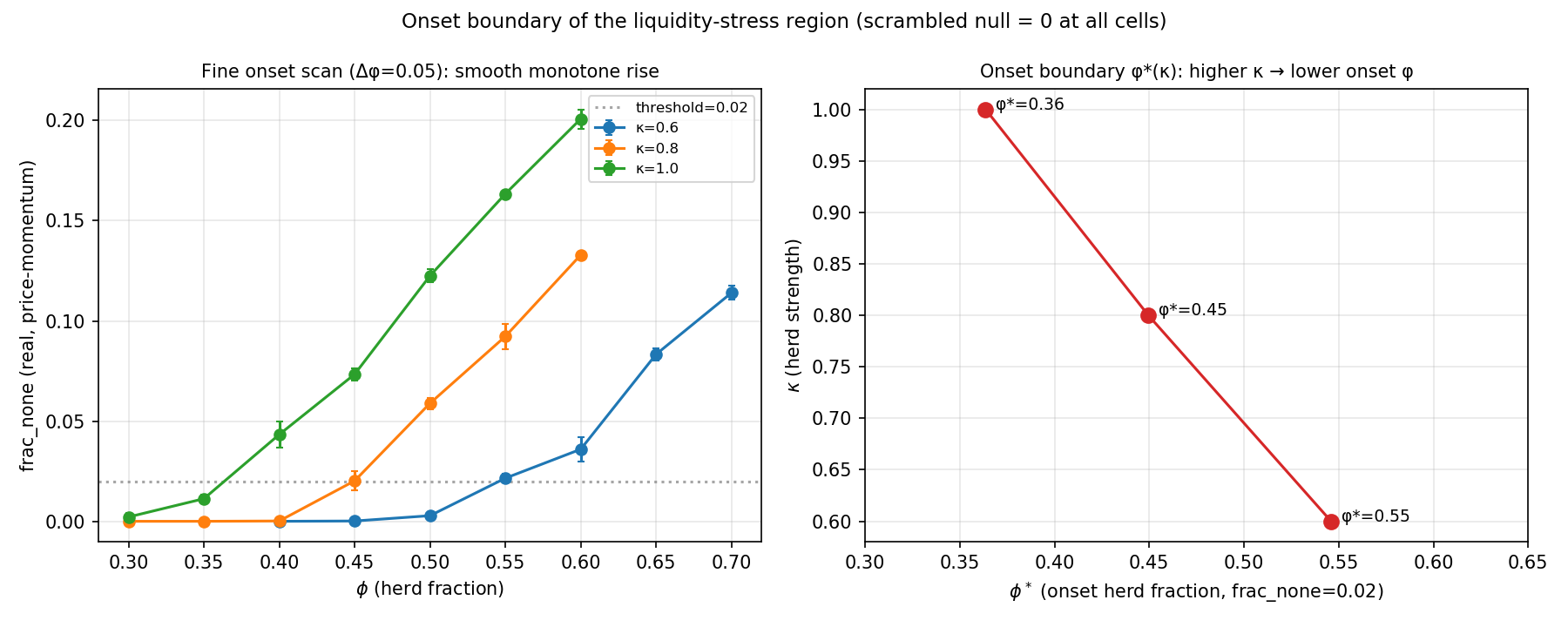}
\caption{Onset boundary of the liquidity-stress region (scrambled null $=0$ at all
cells). \emph{Left:} the fine $\varphi$ scan ($\Delta\varphi=0.05$) at three $\kappa$
rows rises smoothly and monotonically through the $0.02$ threshold, with no jump.
\emph{Right:} the extracted onset boundary $\varphi^*(\kappa)=0.55,0.45,0.36$--higher
herding strength $\kappa$ lowers the onset fraction $\varphi^*$.}
\label{fig:onset}
\end{figure}

A fine $\varphi$ scan ($\Delta\varphi=0.05$) at $\kappa\in\{0.6,0.8,1.0\}$ resolves the
onset of the stress region (Fig.~\ref{fig:onset}, left). Every row rises \emph{smoothly
and monotonically}--per-step increments $\sim0.01$--$0.04$, no jump, no seed
bimodality--with the scrambled null $0.000$ across all $21$ cells. Defining the
onset boundary $\varphi^*(\kappa)$ as the herd fraction at which $\fnone$
first exceeds $0.02$ (linear interpolation) gives a clean, monotone-decreasing curve
(Table~\ref{tab:onset}, Fig.~\ref{fig:onset}, right): $\varphi^*=0.55,0.45,0.36$ at
$\kappa=0.6,0.8,1.0$. \emph{Stronger herding lowers the onset fraction}; the
liquidity-stress region is the wedge above and to the right of this boundary in the
$\varphi\times\kappa$ plane. The same smooth monotone rise holds across all three
$\kappa$ rows.

\begin{table}[htbp]
\centering
\caption{Onset boundary $\varphi^*(\kappa)$--the herd fraction at which
$\fnone$ first exceeds $0.02$. A smooth, monotone-decreasing boundary:
stronger imitation strength lowers the onset fraction.}
\label{tab:onset}
\begin{tabular}{c|ccc}
\toprule
$\kappa$ & 0.6 & 0.8 & 1.0 \\
\midrule
$\varphi^*$ & 0.55 & 0.45 & 0.36 \\
\bottomrule
\end{tabular}
\end{table}

The boundary shape is threshold-robust. The $0.02$ level that defines $\varphi^*$ is a
convention, so we recompute the boundary by linear interpolation at four $\fnone$
thresholds spanning an order of magnitude (Table~\ref{tab:threshold}). The absolute
$\varphi^*$ shifts by about $\pm0.05$--$0.10$ as the threshold moves across
$0.005$--$0.05$, but the boundary is monotone-decreasing in $\kappa$ at every threshold.
The shape a reader should trust is thus threshold-invariant, consistent with the
soft-contour (not tipping-line) framing.

\begin{table}[htbp]
\centering
\caption{Threshold-robustness of the onset boundary $\varphi^*(\kappa)$, recomputed
by linear interpolation at four
$\fnone$ thresholds. The absolute $\varphi^*$ moves $\pm0.05$--$0.10$ over
the range; the monotone-decreasing shape is invariant.}
\label{tab:threshold}
\begin{tabular}{c|ccc}
\toprule
$\fnone$ threshold & $\varphi^*(\kappa{=}0.6)$ & $\varphi^*(\kappa{=}0.8)$ & $\varphi^*(\kappa{=}1.0)$ \\
\midrule
$0.005$              & $0.506$ & $0.412$ & $0.316$ \\
$0.010$              & $0.519$ & $0.425$ & $0.343$ \\
$0.020$ (as reported)& $0.546$ & $0.450$ & $0.364$ \\
$0.050$              & $0.615$ & $0.489$ & $0.411$ \\
\bottomrule
\end{tabular}
\end{table}

Let us comment on the stiff and sloppy directions. Across the swept directions the local
$\fnone$ gradient is dominated by $\varphi$ and $\kappa$ (the onset boundary is entirely a
$\varphi\times\kappa$ curve), while the horizon directions are far sloppier. The window
$w$ barely moves the corner (Section~\ref{sec:horizon}), and $\text{scale}$ matters only
once it is pushed to weaken the signal substantially. The stiff directions are thus
$\{\varphi,\kappa\}$, and $\{w,\text{scale}\}$ are sloppy except in scale's
strong-weakening limit. This is a qualitative reading in the spirit of
\citep{waterfall2006sloppy,bouchaud2024navigating}, short of a formal sensitivity-spectrum
analysis.

\subsection{The mechanism: a robust self-reinforcing component under momentum, comparator-dependent under OFI}
\label{sec:mechanism}

The existence and onset of the dry-up are rule-robust; we now measure its generating
feedback.

Let us measure the reflexive component in a way that is fair to both rules. The reflexive
component of a herding loop is the extra dry-up the closed loop produces over an open-loop
drive of equal forcing, so the drive must genuinely be of equal forcing. Two subtleties
make this non-trivial. The first is that the two rules must be matched in herd-signal
amplitude. The herd side maps its signal through
$p_{\rm buy}=0.5+0.5\kappa\tanh(\mathrm{signal}/\mathrm{scale})$, and the two raw signals
sit at very different points of that nonlinearity. At a shared $\mathrm{scale}=2.0$ the
momentum signal (std $\approx1.8$) is in the tanh's linear regime
($\mathrm{mean}\,|p_{\rm buy}-0.5|\approx0.26$), whereas the OFI signal (std
$\approx21$--$34$) drives it as a near-hard sign function ($\approx0.47$). Comparing them
at a shared scale would compare loops of different directional force, and any contrast
would partly reflect that amplitude gap rather than the loop. We instead match the
closed-loop directional-bias amplitude. Momentum is held at its published
$\mathrm{scale}=2.0$ (the reference target; $\mathrm{mean}\,|p_{\rm
buy}-0.5|=0.2694\pm0.0006$), and OFI's scale is bisected directly in the closed loop until
it reaches the same amplitude, converging at $\mathrm{scale}=68.8$ ($0.2677\pm0.0028$, a
$<1\%$ gap). The matched OFI scale is $\sim34\times$ momentum's, which measures how much
more concentrated near $\pm1$ the raw OFI signal is. It must be found in the closed loop,
because the feedback itself inflates the realised OFI magnitude (an open-loop analytic
match under-guesses at $\mathrm{scale}\approx21$). The second subtlety is that matching the
first moment is not enough. An amplitude match fixes only the mean directional bias of the
drive, not its higher-moment (burst/persistence) structure, and two open-loop drives of
identical mean amplitude but different burstiness can dry the book by very different
amounts. A single open-loop comparator can therefore mislead, so we measure each reflexive
component against several: an open-loop \textsc{shadow} (baseline $\varphi=0$ market flow),
a loop-isolating \textsc{replay} (the closed loop's own signal, applied open-loop), and
clean synthetic telegraph/continuum drives (Section~\ref{sec:continuum}). A component that
agrees across comparators is robust; one that depends on the comparator is not.

\begin{figure}[htbp]
\centering
\includegraphics[width=\textwidth]{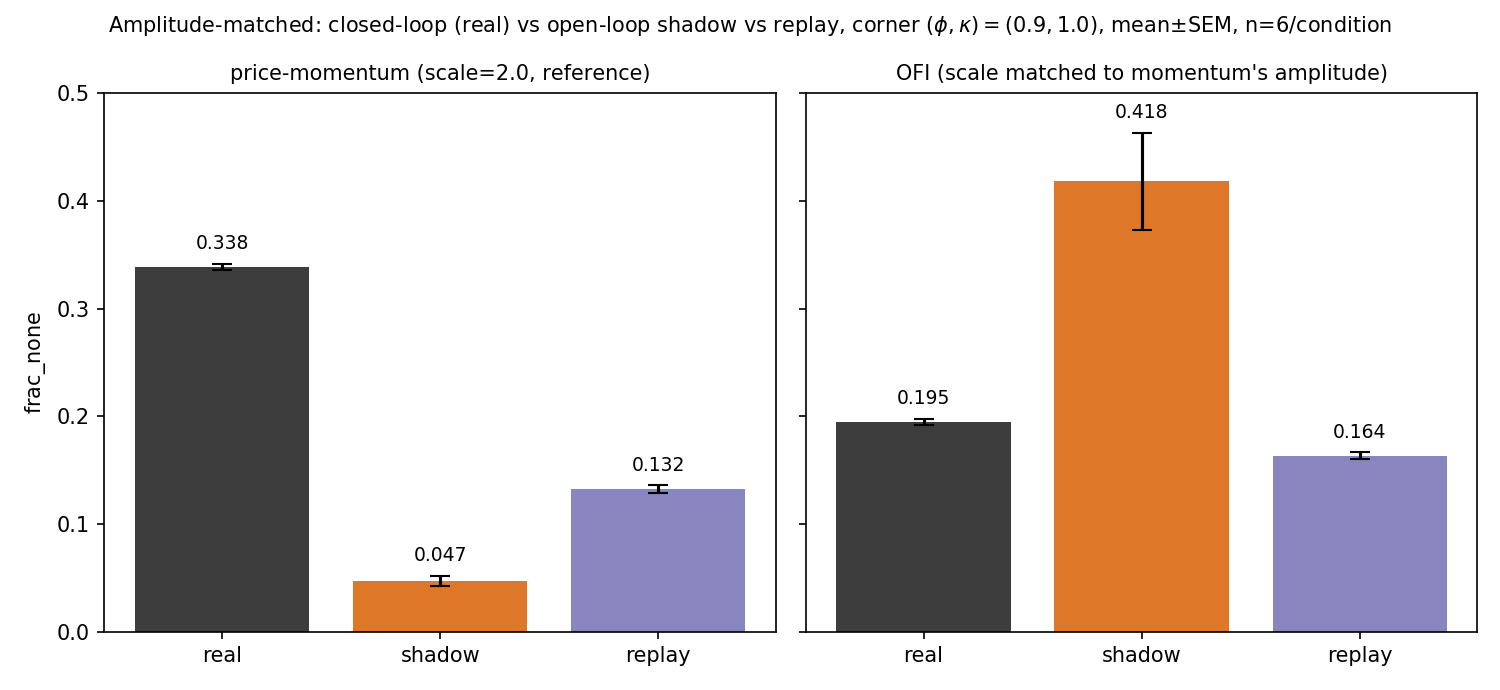}
\caption{Momentum has a robust self-reinforcing reflexive component; OFI's is
comparator-dependent ($\approx0$), at matched herd-signal amplitude
($\mathrm{mean}\,|p_{\rm buy}-0.5|\approx0.269$). \emph{Left (momentum):} the closed loop
(\textsc{real}, $0.338$) manufactures $\sim7\times$ more dry-up than \emph{every}
equally-forceful open-loop drive (\textsc{shadow}, $0.047$; \textsc{replay},
$0.132$)--strong, comparator-robust self-reinforcement. \emph{Right (OFI):} the
baseline-flow
\textsc{shadow} ($0.418$) exceeds \textsc{real} ($0.195$), but the loop-isolating
\textsc{replay} ($0.164$) sits \emph{below} \textsc{real}, so the OFI component is not
robustly signed--the shadow is an outlier driven by higher-moment (burst) structure,
not by absence of the loop. Error bars are seed SEM (the OFI shadow bar is the noisier,
driving-series-dependent one).}
\label{fig:shadow}
\end{figure}

Let us now compare the real loop with the amplitude-matched shadow and replay. At matched
amplitude, the corner decomposition (Table~\ref{tab:shadow}, Fig.~\ref{fig:shadow}) is
commensurable across rules. Both the shadow and replay legs use one consistent burn-in
convention (the full recorded closed-loop $p_{\rm buy}$ series, no neutral padding) for
both rules.

\begin{table}[htbp]
\centering
\caption{Corner $(0.9,1.0)$ $\fnone$ under the \emph{amplitude-matched}
shadow / replay decomposition (mean$\pm$SEM, $\ge6$ seeds/condition; shadow uses
$\ge3$ independent driving series). Momentum's dry-up is \emph{far above both} its
open-loop shadow and its replay (strong, comparator-robust self-reinforcement). OFI's
real value ($0.195$) sits \emph{below} the baseline-flow shadow ($0.418$) but
\emph{above} the loop-isolating replay ($0.164$), so its reflexive component flips sign
with the comparator--it has no robust sign (see text).}
\label{tab:shadow}
\begin{tabular}{l|cc}
\toprule
condition & price-momentum (scale $2.0$) & OFI (scale $68.8$, matched) \\
\midrule
\textsc{real} (closed loop) & $\mathbf{0.338\pm0.003}$ & $\mathbf{0.195\pm0.003}$ \\
\textsc{shadow} (open-loop, matched amplitude) & $0.047\pm0.004$ & $0.418\pm0.045$ \\
\textsc{replay} (closed-loop signal, open loop) & $0.132\pm0.004$ & $0.164\pm0.003$ \\
\bottomrule
\end{tabular}
\end{table}

The reflexive component is \textsc{real} minus an amplitude-matched open-loop drive: the
extra dry-up the closed loop produces over an equally-forceful open-loop injection.
Table~\ref{tab:reflexive} evaluates it against each of the three matched comparators. The
two rules behave differently. Momentum's component is large and positive against every
comparator--$+0.29$ vs the shadow, $+0.21$ vs the replay, and $\approx+0.34$ vs the
persistence-matched synthetic telegraph--so its sign is not an artefact of any one
baseline. OFI's component, by contrast, flips with the comparator: it is $-0.22$ against
the baseline-flow shadow but $+0.03$ against the loop-isolating replay and $\approx0$
($+0.05$) against the synthetic telegraph and the continuum grid. Momentum has a robust
self-reinforcing reflexive component; OFI's has no robust sign.

\begin{table}[htbp]
\centering
\caption{Reflexive component (\textsc{real} $-$ open-loop drive) at the corner, at matched
herd-signal amplitude ($\mathrm{mean}\,|p_{\rm buy}-0.5|\approx0.269$), against each of
three comparators. \textsc{real} is $0.338$ (momentum) / $0.195$ (OFI). Momentum's
component is large and positive against \emph{all} comparators (comparator-robust
self-reinforcement). OFI's changes sign with the comparator: strongly negative only
against the baseline-flow shadow--whose value ($0.418$) is a $\sim2\times$ outlier
among matched-amplitude drives, driven by unmatched burst structure--and $\approx0$
against the loop-isolating replay and the clean synthetic drives, so it has no robust
sign. (The shadow leg draws on only $3$ independent driving series, so its
seed-level SEM understates the true uncertainty; clustering on driving series widens the
OFI--shadow interval substantially. The \emph{sign} against the shadow is stable, but the
comparator disagreement, not the shadow's nominal significance, is the key point.)}
\label{tab:reflexive}
\begin{tabular}{l|cc}
\toprule
open-loop comparator (matched amplitude) & momentum component & OFI component \\
\midrule
open-loop \textsc{shadow} (baseline $\varphi{=}0$ flow) & $\mathbf{+0.291\pm0.005}$ & $-0.223\pm0.045$ \\
loop-isolating \textsc{replay} (own signal)             & $+0.206$ & $+0.031$ \\
persistence-matched synthetic telegraph / grid          & $\approx+0.34$ & $\approx0$ ($+0.05$) \\
\bottomrule
\end{tabular}
\end{table}

We read this as a robust momentum component and an ambiguous OFI one. Under price momentum
the closed loop manufactures about seven times more dry-up than an amplitude-matched
exogenous drive, and does so against every open-loop comparator: shadow ($0.338$ vs
$0.047$), replay ($0.338$ vs $0.132$), and every synthetic-telegraph cell in
Section~\ref{sec:continuum}. The positive sign also holds at a second matched amplitude
($\approx+0.19$ at $\mathrm{mean}\,|p_{\rm buy}-0.5|\approx0.15$ in an independent
reproduction). This is genuine, strong, comparator-robust self-reinforcing feedback: the
herd reads the mid it is itself moving, so buying begets buying, and the loop generates
more one-sided pressure than any equally-forceful open-loop stream.

Under OFI the picture is different. The reflexive component is comparator-dependent and
$\approx0$, so it has no robust sign. The $-0.223$ arises only against the baseline-flow
shadow ($0.418$). At the same matched amplitude the loop-isolating replay gives
$0.195-0.164=+0.031$, a persistence-matched telegraph gives $0.17$--$0.21$, and the
continuum grid's nearest cell gives $0.143$, all $\approx$ the real value ($0.195$). Three
of the four comparators therefore put the OFI component at $\approx0$; only the
baseline-flow shadow, whose value ($0.418$) is a $\sim2\times$ outlier among
matched-amplitude drives, makes it strongly negative. This is the higher-moment effect
anticipated above. Matching the herd-side first-moment amplitude did not match the
burst/persistence structure of the drive, and the baseline-market flow that generates the
shadow is far burstier than the closed loop's own signal at equal mean amplitude (the same
gap by which the OFI shadow sits $\sim3\times$ above the persistence-matched telegraph,
Section~\ref{sec:continuum}). The OFI shadow's excess is thus a residual signal-source
property of the drive, not evidence that the loop self-limits. The outcome is asymmetric:
momentum has a robust, comparator-robust self-reinforcing reflexive component, while OFI's
reflexive component is $\approx0$ and comparator-dependent, with no robust sign. The two
are not symmetric opposite-signed twins. The most that can be said for OFI is narrower and
about signal generation: the closed loop produces a signal that dries the book less than an
equally-forceful stream of exogenous baseline flow, which is a property of how the loop
generates its own OFI signal, not a self-limiting reflexive component.

\subsection{No cross-market contagion (a corollary of reflexivity)}
\label{sec:contagion}
A self-referential mechanism makes a sharp prediction about \emph{coupled} markets: because the
stress is manufactured by a market acting on its \emph{own} momentum, it should be intrinsically
\emph{local} and not become contagious through a purely behavioural link. A companion two-market
analysis tests exactly this. Two copies of the model are coupled by \emph{cross-market herding}:
only a scalar momentum signal crosses between the books--no order-flow leakage, so A's orders,
trades and book never touch B--with coupling strength $c$. The coupling is foundation-gated:
at $c{=}0$ the two-market wrapper reduces exactly to two independent markets, each reproducing the
single-market corner ($\fnone\approx0.34$) and the exact-zero own-scrambled null.

\begin{figure}[t]
\centering
\includegraphics[width=0.64\textwidth]{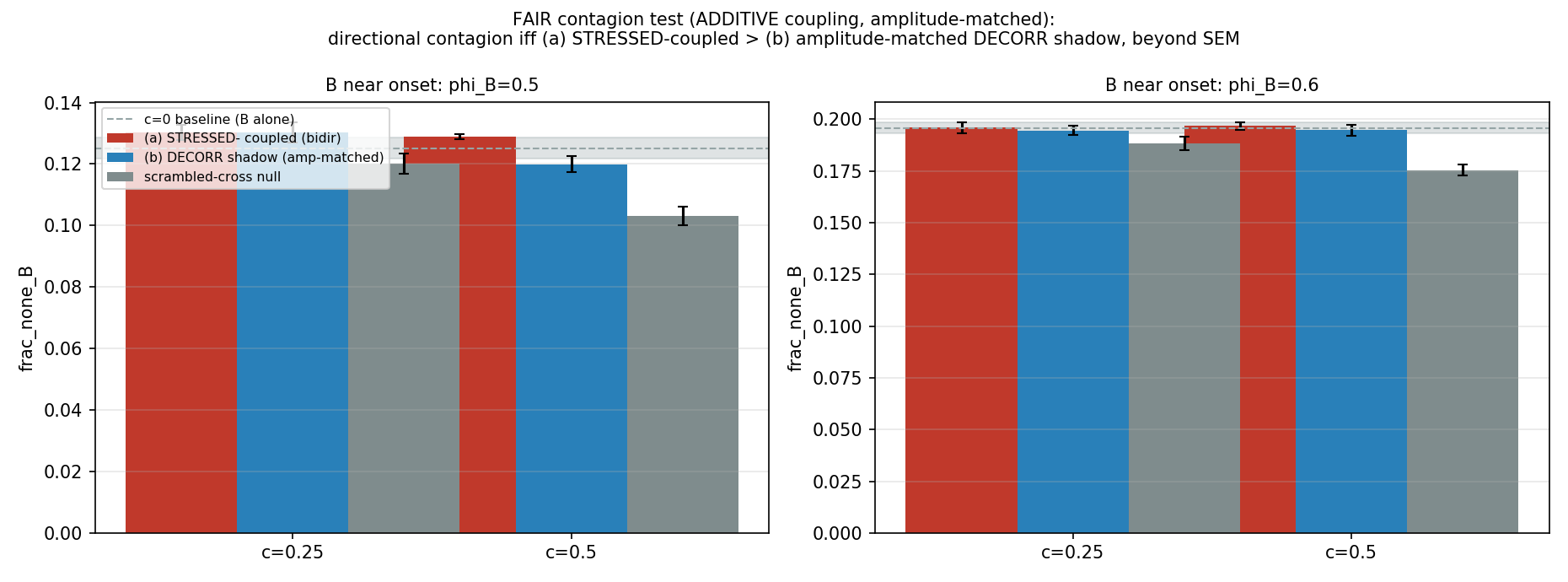}
\caption{No cross-market contagion (decisive fair test). With B near its own onset and A's push
\emph{added} at matched amplitude, coupling to a genuinely stressed A overlaps an amplitude-matched
decorrelated shadow and the $c{=}0$ baseline: B's stress tracks injected amplitude, not whether
the neighbour is stressed--no directional contagion, as the self-referential mechanism predicts.
}
\label{fig:contagion}
\end{figure}

Let us set up the decisive fair test. We hold B near its own onset
($\varphi_B\in\{0.5,0.6\}$), where a contagious nudge would be most visible, and add A's
push to B's own signal so that B keeps its full own reflexivity. We then compare coupling B
to a genuinely stressed, co-evolving A against coupling to an amplitude-matched
decorrelated recording of equal injected strength (matched to $4$ decimals; $8$ seeds,
$\pm$SEM). The two are statistically indistinguishable: at $\varphi_B{=}0.6$, coupling to a
stressed A gives
$\phi_{\varnothing,B}=0.196\pm0.003$ versus $0.195\pm0.002$ for the decorrelated shadow, both
astride the $c{=}0$ baseline of $0.196$ (Figure~\ref{fig:contagion}).
The null is robust to doubling the coupling strength (to $c{=}2.0$) and to the coupling form.

There is therefore \emph{no directional, state-dependent contagion}: B's stress is set by the
injected signal \emph{amplitude}, not by whether the neighbour is genuinely stressed. We state
this precisely rather than as a blanket ``no spillover.'' A different signal-only
channel--herd-intensity modulation, in which B's agents herd more often when A is
turbulent--can lift a quiet, deeply sub-critical B's stress from $\approx0$ to
$\approx0.198$, but again by amplitude (live stressed-A $\approx$ an equal-amplitude
decorrelated shadow), not because the neighbour is stressed.

Let us say why this is a corollary. The single-market crossover is genuinely
self-referential (Section~\ref{sec:mechanism}): its stress is manufactured within a book,
buying begetting buying. Cross-market herding injects an external, open-loop signal into B,
and an external signal--however stressed its source--cannot instantiate B's own internal
loop. What the fair test holds amplitude-matched is precisely the ``push,'' isolating the
fact that what B lacks is not a stronger drive but a drive that is its own. The
no-contagion result is thus a prediction of the reflexive mechanism rather than an
independent finding. This
signal-only herding channel differs from portfolio-network order-book contagion
\citep{paulin2018flashcrash} and from continuum Hopf-bifurcation contagion in multi-asset flow
models \citep{cavani2026fused}, and from cross-asset learning \citep{cespa2014illiquidity},
fire-sale order flow \citep{contwagalath2013running} and empirical cross-impact
\citep{benzaquen2017crossimpact}--all of which move balance-sheet or order-flow content
between markets, whereas our channel moves only a scalar signal and transmits no stress.

\subsection{The dry-up magnitude follows an amplitude\texorpdfstring{$\times$}{x}persistence continuum}
\label{sec:continuum}
\begin{figure}[htbp]
\centering
\includegraphics[width=0.8\textwidth]{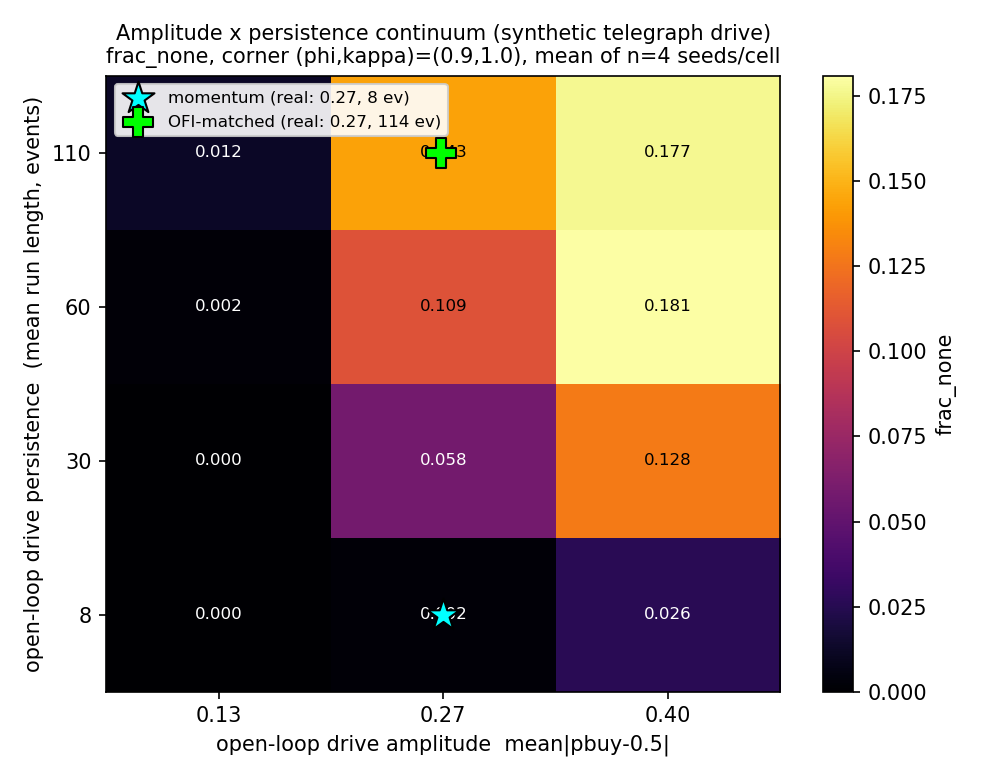}
\caption{Amplitude $\times$ persistence continuum of the dry-up under a synthetic
telegraph drive at the corner. $\fnone$ increases smoothly and monotonically
along both axes, with no threshold. Momentum's and matched-OFI's real operating points are
overlaid; the OFI operating point's actual open-loop shadow ($0.418$) sits well above the
nearest synthetic cell ($0.143$), showing the continuum is qualitatively but not
quantitatively complete (see caveat).}
\label{fig:continuum}
\end{figure}

Is the dry-up's \emph{magnitude} a smooth function of an exogenous drive's statistics, or
does it switch on categorically? Replacing the herd side at the corner with a synthetic
two-state (telegraph) drive whose amplitude and persistence are set \emph{independently}
(Section~\ref{sec:iter2exp}) answers this cleanly: $\fnone$ rises
\emph{smoothly with both knobs}, with no threshold and no categorical jump
(Table~\ref{tab:continuum}, Fig.~\ref{fig:continuum}). The rise is monotone in both
directions across the grid, with a single within-SEM exception--at the highest
amplitude ($0.40$) the value plateaus and dips marginally from persistence $60$ to $110$
($0.181\to0.177$), a saturation effect, not a reversal of the trend.

\begin{table}[htbp]
\centering
\caption{$\fnone$ under a synthetic two-state exogenous drive at the corner,
amplitude $\times$ persistence grid ($4$ seeds/cell). Smooth and monotone in both
directions--a genuine continuum--up to a within-SEM plateau at the highest
amplitude ($0.40$, persistence $60\to110$: $0.181\to0.177$). The extreme rows are
labelled by which real rule's raw-signal persistence they bracket.}
\label{tab:continuum}
\begin{tabular}{l|ccc}
\toprule
persistence $\backslash$ amplitude & $0.13$ & $0.269$ & $0.40$ \\
\midrule
$8$ ev (momentum-like)  & $0.000$ & $0.002$ & $0.026$ \\
$30$ ev                 & $0.000$ & $0.058$ & $0.128$ \\
$60$ ev                 & $0.002$ & $0.109$ & $0.181$ \\
$110$ ev (OFI-like)     & $0.012$ & $0.143$ & $0.177$ \\
\bottomrule
\end{tabular}
\end{table}

One caveat applies to the continuum. It is real and load-bearing, since both amplitude and
persistence matter smoothly, with no on/off threshold. But it is a two-parameter
(amplitude, persistence) idealisation and does not fully account for the real rules' own
shadow values. At matched coordinates, momentum's actual shadow ($0.047$) is somewhat above
the synthetic grid's nearest prediction ($0.002$), and OFI's actual shadow ($0.418$) is
$\sim3\times$ above the nearest synthetic cell ($0.143$). The real OFI signal is not a
symmetric telegraph process: its higher-moment / burst structure matters beyond mean
amplitude and mean run-length. This is the same signal-statistics gap that makes the OFI
reflexive component comparator-dependent in Section~\ref{sec:mechanism}. The baseline-flow
shadow's $\sim2$--$3\times$ outlier value is a burst-structure effect, not an ``absence of
the loop'' effect, so $\textsc{real}-\textsc{shadow}$ is not a clean measure of the OFI
loop's own feedback (whereas replay, the telegraph, and this grid, all closer in
higher-moment structure, agree the component is $\approx0$). The continuum settles the
qualitative question (smooth, not categorical), but a full quantitative account of the OFI
shadow needs the raw signal's higher-order statistics.

\subsection{The RMS-mispricing gradient is a placement/dilution artefact}
\label{sec:artefact}

$\mathrm{RMS}|{\rm mid}-{\rm fund}|$ does rise toward high $\varphi$, but its
$\kappa$-dependence shows it is \emph{not} the imitation channel (Table~\ref{tab:rms}).
The gradient is \emph{largest at $\kappa=0$}--reaching $\approx6$ at
$(\varphi,\kappa)=(0.9,0)$, where directional herding is off--and does \emph{not}
increase with $\kappa$; if anything it falls. At $\kappa=0$ a ``herd'' order is just a
random-side, mid-anchored order that dilutes the fundamental-anchored restoring force:
the known placement/composition effect, present in the null too.

\begin{table}[htbp]
\centering
\caption{$\mathrm{RMS}|{\rm mid}-{\rm fund}|$, REAL condition, mean over $5$ seeds.
Rows: $\kappa$; columns: $\varphi$. The mispricing gradient is largest along the
$\kappa=0$ row (herding \emph{off}) and is flat/low along $\kappa=1.0$ despite the
strongest imitation--the opposite of what an imitation-driven Dark Corner would
show.}
\label{tab:rms}
\begin{tabular}{c|ccccccc}
\toprule
$\kappa \backslash\, \varphi$ & 0.0 & 0.15 & 0.3 & 0.45 & 0.6 & 0.75 & 0.9 \\
\midrule
1.0 & 2.77 & 1.80 & 2.08 & 2.51 & 2.37 & 2.48 & 2.58 \\
0.8 & 3.70 & 1.96 & 2.07 & 2.63 & 2.45 & 2.63 & 2.67 \\
0.6 & 3.06 & 2.69 & 2.18 & 2.55 & 2.89 & 2.89 & 2.79 \\
0.4 & 3.08 & 2.40 & 2.41 & 2.61 & 2.79 & 3.41 & 3.36 \\
0.2 & 2.89 & 2.36 & 2.57 & 2.08 & 3.19 & 2.68 & 5.44 \\
0.0 & 2.57 & 2.39 & 3.10 & 3.37 & 3.49 & 4.94 & \textbf{5.97} \\
\bottomrule
\end{tabular}
\end{table}

\begin{figure}[htbp]
\centering
\includegraphics[width=0.75\textwidth]{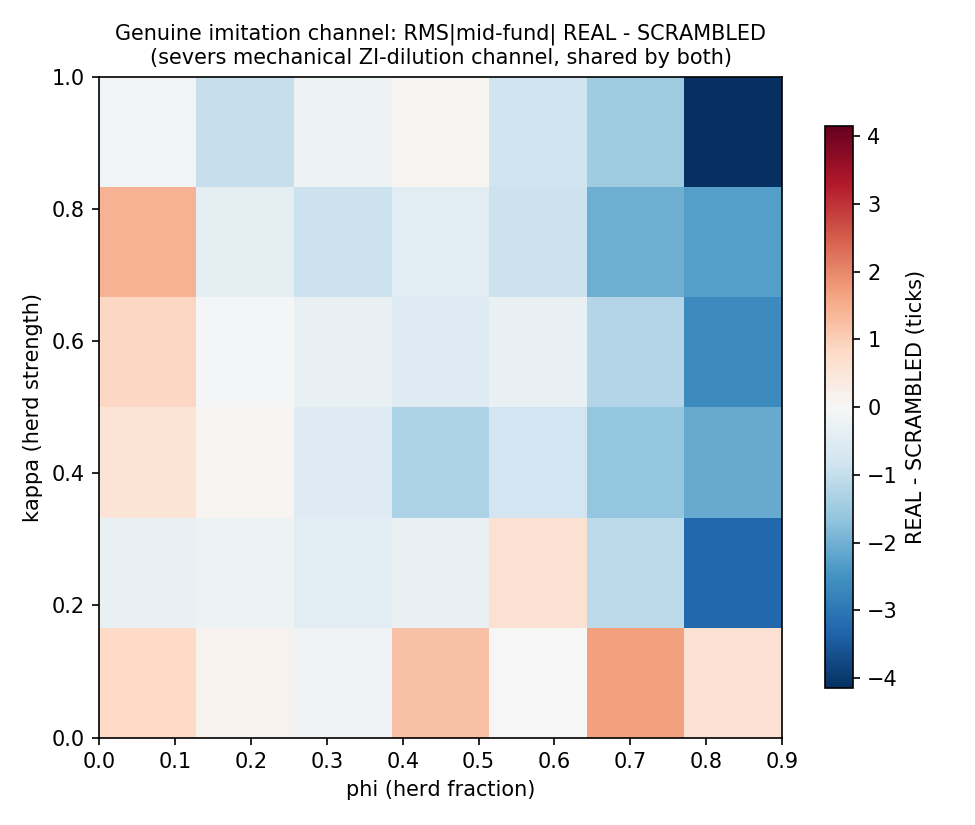}
\caption{Channel decomposition, REAL $-$ SCRAMBLED for RMS mispricing. The imitation
channel does not add mispricing over the placement/dilution baseline captured by the
null; consistent with the reading that the RMS gradient is a placement artefact.}
\label{fig:channel}
\end{figure}

The $\kappa=1.0$ cut (Fig.~\ref{fig:cut}, left panel) shows this directly: the
\emph{scrambled} curve rises to $\approx6.7$ at $\varphi=0.9$ while the \emph{real}
curve stays flat at $\approx2.5$. Real directional imitation suppresses mispricing: it
pins the price, at the cost of the liquidity dry-up documented above. The mispricing rise
therefore belongs to the random-side placement/dilution channel, not to imitation. The
band cut agrees: with the same continuous-reference RMS definition as the main sweep
(corner RMS $2.62$ at band $4$), real RMS at $\kappa=1.0$ stays low across bands, never
exploding. We therefore do not present RMS-mispricing growth as a Dark Corner. The channel
decomposition (real $-$ null,
Fig.~\ref{fig:channel}) shows a negative-to-neutral RMS channel, consistent with this
reading.

\subsection{Other order parameters show no tipping structure}
\begin{figure}[htbp]
\centering
\includegraphics[width=\textwidth]{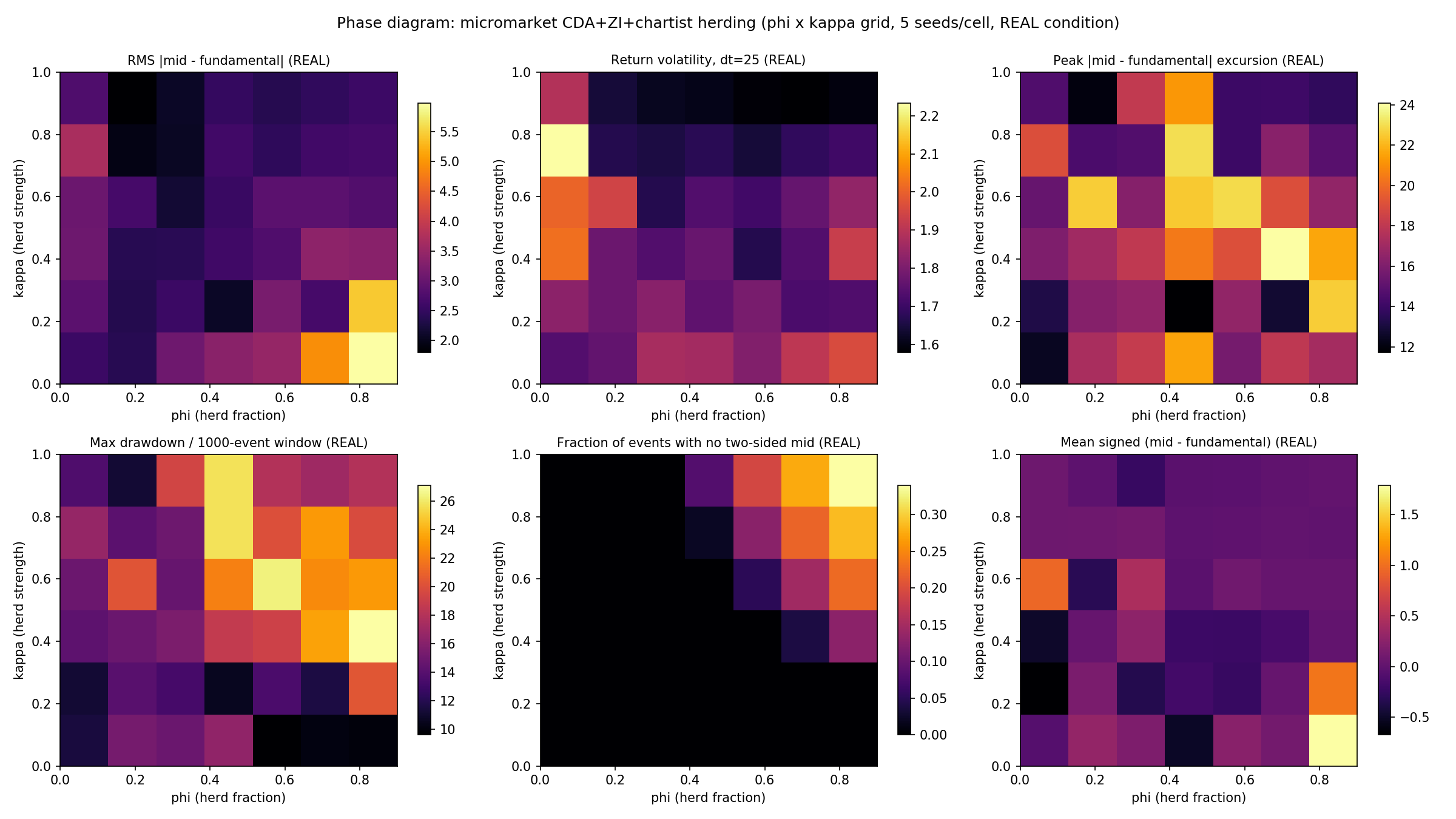}
\caption{All six order parameters across the $\varphi\times\kappa$ plane (REAL
condition). Only $\fnone$ (the liquidity dry-up) shows a clean,
localised region; the mispricing, excursion, drawdown and volatility panels show no
tipping structure.}
\label{fig:heatmaps}
\end{figure}

For completeness, Fig.~\ref{fig:heatmaps} shows all six order parameters (real
condition). Beyond $\fnone$, none shows tipping structure.
\begin{itemize}
  \item Peak absolute deviation and maximum drawdown are noisy ($\approx12$--$24$ and
    $\approx10$--$27$ respectively) with no clean $\varphi\times\kappa$ structure, and
    real $\approx$ null across the plane. There is no excursion or drawdown tipping line.
  \item Return volatility ($\mathrm{d}t=25$) is flat to mildly decreasing with $\kappa$
    (the $\kappa=1.0$ row $\approx1.6$--$1.9$ vs.\ baseline $\approx1.7$--$2.0$),
    consistent with the sibling finding that this OFI/momentum chartist damps tick
    volatility. There is no volatility runaway.
\end{itemize}

\section{Discussion}

\subsection{A liquidity phenomenon, not a price phenomenon}
The data support a robust observable with a rule-specific reflexive signature. When a
large fraction of strongly imitating agents ($\varphi$ and $\kappa$ both high) act on a
shared directional signal, their orders become correlated. They then repeatedly consume
one side of the finite book faster than fundamental-anchored ZI liquidity can replenish
it, so the book goes one-sided a substantial fraction of the time ($\fnone\to0.34$ under
momentum, $\to0.23$ under OFI). The price does not run away. Directional imitation, by
pinning the mid to the consensus, actually lowers RMS mispricing relative to the placement
baseline. The stress is therefore in liquidity (book sidedness), not in price (no bubble,
no crash). This is why $\fnone$, and not any mispricing or drawdown metric, is the order
parameter that separates the stressed from the quiescent regime.

The scrambled-sign null is what licenses the causal reading. The null preserves order-type
mix and ZI dilution but destroys directional correlation, and $\fnone=0$ across the entire
null plane, under momentum and under OFI. We therefore attribute the dry-up to
directionally correlated imitation rather than to a mechanical reduction of order flow.
This is the distinction between a genuine emergent collective effect and a composition
artefact \citep{filimonov2012reflexivity}. The same discipline applies to the
RMS-mispricing analysis in reverse: the mispricing gradient fails the same test (it is
largest with imitation off and is suppressed by imitation), so we decline to call it a
Dark Corner.

\subsection{The reflexive component: robust under momentum, unsigned under OFI}
The robustness battery does two things at once. It strengthens the existence claim, since
the dry-up survives a change of behavioural rule (OFI), a $16\times$ change of herding
horizon, and admits a clean, threshold-robust onset boundary $\varphi^*(\kappa)$. It also
establishes a robust, self-reinforcing reflexive component under price momentum. Isolating
that component fairly is what the amplitude-matched, multi-comparator protocol
(Section~\ref{sec:mechanism}) is for. The two rules' raw signals sit at very different
points of the response nonlinearity, so they must be driven at a matched herd-signal
amplitude ($\mathrm{mean}\,|p_{\rm buy}-0.5|\approx0.269$) before their loops are compared.
That first-moment match still leaves higher-moment (burst) structure unmatched, so each
loop must be checked against several open-loop comparators. Under price momentum the
reflexive component is then large, positive, and comparator-robust: $+0.29$ against the
open-loop shadow, $+0.21$ against the loop-isolating replay, far above every
synthetic-telegraph cell, and positive again at a second matched amplitude. The
reflexivity that \citet{filimonov2012reflexivity} frame as the route to endogenous
instability is thus genuinely and strongly self-reinforcing under momentum. The herd reads
a price it is itself moving, so buying begets buying and the closed loop manufactures
$\sim7\times$ more dry-up than any equally-forceful exogenous drive.

Under OFI, by contrast, the reflexive component is $\approx0$ and comparator-dependent,
with no robust sign. The negative value ($-0.223$) holds only against the single
baseline-flow shadow ($0.418$). At the same matched amplitude the loop-isolating replay
gives $+0.03$, and a persistence-matched telegraph ($0.17$--$0.21$) and the continuum grid
($0.143$) both agree with the real value ($0.195$). The baseline-flow shadow is a
$\sim2\times$ outlier among matched-amplitude comparators because it is far burstier than
the closed loop's own signal at equal mean amplitude, the same signal-statistics gap by
which it sits $\sim3\times$ above the persistence-matched telegraph. The OFI shadow's
excess thus reflects a property of the drive, not a self-limiting loop. The defensible
narrow statement is only that the closed loop generates a signal that dries the book less
than an equally-forceful stream of exogenous baseline flow. The conclusion is asymmetric:
one robust observable with a genuine self-reinforcing reflexive component under momentum,
and an OFI component that is not robustly signed. The dry-up's magnitude, separately, is a
smooth function of an exogenous drive's amplitude and persistence (a genuine continuum,
monotone up to a within-SEM plateau at the highest amplitude), though a two-parameter
telegraph idealisation does not fully reproduce the real OFI shadow (which sits
$\sim3\times$ higher, the same higher-order signal structure at work).

\subsection{Relation to the closest prior work}
Gao et al.\ \citep{gao2022flashcrash} study crash pre-emption in an order-book ABM where
the crash is triggered by an exogenous shock or bubble-collapse mechanism calibrated to the
2010 flash crash. Our result differs on both axes. First, the stress here is endogenous,
emerging purely from the interaction of directional imitation with a finite book and with
no exogenous trigger. Second, it is subjected to an explicit null (scrambled-sign),
robustness (rule-change, horizon), and artefact (band, numerical-stability) discipline,
plus an amplitude-matched mechanism decomposition, and the outcome is a smooth
liquidity-stress crossover rather than a discontinuous crash. We make no claim of a sharp
tipping line, which is what distinguishes this from crash-pre-emption framings and from the
sharp Dark Corners of the macroeconomic lineage
\citep{gualdi2015tipping,gualdi2016darkcorners}.

The self-referential mechanism has a corollary: it predicts that the stress is
intrinsically local. A companion two-market analysis (Section~\ref{sec:contagion}) confirms
this. Coupling a near-onset market to a genuinely stressed neighbour through cross-market
herding is statistically indistinguishable from coupling it to an amplitude-matched
decorrelated signal. An external open-loop signal cannot instantiate a market's internal
loop, so no directional contagion crosses. This is predicted by, and consistent with, the
reflexive reading rather than an independent finding. 

\subsection{Limitations}
\label{sec:limitations}
\begin{itemize}
  \item Momentum's reflexive component is robust; OFI's is not. Momentum's positive,
    self-reinforcing component reproduces against every comparator (shadow, replay,
    synthetic telegraph) and stays positive ($\approx+0.19$) at a second amplitude on
    independent reproduction. OFI's component is comparator-dependent ($-0.22$ against the
    baseline-flow shadow, but $+0.03$ against the loop-isolating replay and $\approx0$
    against a persistence-matched telegraph and the continuum grid), so we make no sign
    claim for it. Which sign, if any, a plausible chartist heuristic carries is therefore
    not mapped even for these two rules: one is robust, one is ambiguous.
  \item The OFI reflexive component reflects a residual signal-statistics confound, and its
    against-shadow significance is pseudo-replicated. Matching the herd side's first-moment
    amplitude ($\mathrm{mean}\,|p_{\rm buy}-0.5|\approx0.269$) does not match its
    higher-moment (burst/persistence) structure; the baseline-flow shadow's value ($0.418$)
    is a $\sim2\times$ outlier versus every other matched-amplitude open-loop comparator,
    driven by that unmatched structure. The nominal OFI--shadow $z$ is correspondingly
    optimistic: the shadow leg has only $3$ independent driving series (per-series means
    span a wide range), so treating the $6$ seeds as independent inflates significance, and
    clustering on driving series widens the interval substantially. In summary, the OFI
    reflexive component is $\approx0$ against clean comparators, and negative and noisy only
    against baseline flow.
  \item The matched target amplitude ($0.269$) is a choice, not a law. It is momentum's own
    default-scale value, and we matched OFI to it (the most natural single reference) rather
    than to some third target. For momentum the positive sign is amplitude-stable (an
    independent reproduction at $\approx0.15$ still gives $\approx+0.19$); for OFI the
    amplitude choice is moot, since no comparator-robust sign survives at $0.269$ in the
    first place.
  \item The continuum is qualitative, not a quantitative model of the real shadows. A
    two-parameter telegraph idealisation reproduces the ``smooth in both knobs, no
    threshold'' structure but underpredicts the real OFI shadow by $\sim3\times$; the raw
    signals' higher moments are not characterised here.
  \item The effect remains a smooth crossover, not a sharp Dark Corner. The fine onset scans
    ($\Delta\varphi=0.05$, three $\kappa$ rows) show smooth monotone rises with no
    discontinuity; no runaway, crash, or order-parameter jump was found. The
    $\varphi^*(\kappa)$ boundary is a soft onset contour, not a tipping line, and its
    monotone shape is invariant across $\fnone$ thresholds $0.005$--$0.05$.
  \item $\fnone$ is still rising at the corner $(0.9,1.0)$ for both signals. The map does
    not reach a plateau, so behaviour beyond $\varphi=0.9$ is uncharacterised. We
    deliberately did not push into $\varphi\to1$, which removes ZI liquidity provision
    almost entirely, a degenerate limit.
  \item The replay burn-in convention matters. Using the full recorded real signal from
    event $0$ rather than a neutral warm-up roughly halves momentum's replay $\fnone$
    ($0.132$ vs $\sim0.33$), i.e.\ $20{,}000$ events do not always erase path-dependence
    when the imposed drive is strongly self-reinforcing early on. This caveat is on the
    replay number only; the reflexive-component headline does not use replay.
  \item The null seed counts are reduced in the robustness sweeps ($2$ scrambled for the OFI
    grid, $1$ per horizon/onset cell). This is justified: the null's $\approx0$ value is
    established at higher $n$ in the main grid, and the robustness sweeps need only confirm
    it stays $\approx0$, which it does (exactly $0$ in every such condition).
  \item The study covers the CDA only, two herding rules, and one placement kernel. There is
    no market-design (CDA vs.\ frequent-batch-auction) comparison, and cancellation-rate and
    other placement rules were not swept.
  \item Compute is a binding constraint. The low-$\varphi$/scrambled runs stay dense and
    dominate the cost of the base sweep, which bounds the seed counts feasible in the
    robustness battery.
\end{itemize}

\subsection{Future work}
The battery here closes several natural checks: the fine onset scan (analyticity), the
herding-horizon sweep, the rule-robustness test, the amplitude-matched mechanism
decomposition that established momentum's robust self-reinforcing component and exposed the
OFI component as comparator-dependent, and the amplitude$\times$persistence continuum. Four
directions remain open. The first is a comparator that matches the drive's higher-moment
(burst/persistence) structure, not just its first moment, so the OFI reflexive component
can be resolved cleanly rather than left comparator-dependent, together with mapping more
behavioural rules to see how common a robust self-reinforcing component (momentum-like) is
versus an ambiguous one (OFI-like). The second is characterising the raw signals'
higher-order statistics so the continuum can account for the real shadows quantitatively,
not just qualitatively. The third is a formal sloppy/stiff parameter-direction
(sensitivity-spectrum) analysis once more than two axes are in play, following the
manifesto's high-dimensional programme \citep{waterfall2006sloppy,bouchaud2024navigating},
of which Section~\ref{sec:onset} gives only a qualitative reading. The fourth is a
market-design comparison (CDA vs.\ frequent-batch-auction) to test whether batching
resolves or moves the OFI reflexive component.

\section{Conclusion}

We applied Bouchaud's phase-diagram / Dark-Corners methodology
\citep{bouchaud2024navigating} to an order-book microstructure ABM. To our knowledge this is
the first time the full discipline---a two-dimensional phase diagram plus an explicit null,
artefact and robustness battery---has been carried into an order-book microstructure vehicle
with herding and liquidity as its control axes. This complements, rather than repeats, the
single-parameter order-book liquidity-crisis transition of \citet{fosset2020liquiditycrises}:
where they find a genuine critical point, we find a smooth crossover. We mapped
the two-dimensional $\varphi\times\kappa$ (herd-fraction $\times$ herd-strength) phase
diagram, applied an explicit null and artefact battery, and then followed the manifesto one
step further, to test the robustness of the resulting feature and isolate its mechanism.

The positive result is a genuine, emergent, null-verified, imitation-specific
liquidity-stress crossover in the high-herding corner. The fraction of one-sided-book
events rises to $\approx0.34$ at $(\varphi,\kappa)=(0.9,1.0)$, requires both high $\varphi$
and $\kappa>0$, is zero across all $42$ scrambled-sign null cells, and is neither numerical
nor degenerate. The robustness battery makes this a characterised phenomenon rather than a
single-specification observation. The dry-up is rule-robust (it reappears under the OFI
herding signal, with the null still identically zero), horizon-robust
($\fnone\approx0.32$--$0.35$ across a $16\times$ range of momentum window), and has a clean,
monotone onset boundary $\varphi^*(\kappa)=\{0.55,0.45,0.36\}$ at $\kappa=\{0.6,0.8,1.0\}$.

The same battery also characterises the mechanism. We compare the two rules at a matched
herd-signal amplitude, and against several open-loop comparators, to guard against
unmatched higher-moment structure. Under price momentum this establishes a robust,
self-reinforcing reflexive component: $+0.29$, positive against every comparator (shadow,
replay, synthetic telegraph) and at a second amplitude. The closed loop makes $\sim7\times$
more dry-up than any equally-forceful exogenous drive, so buying begets buying. Under OFI,
by contrast, the reflexive component is $\approx0$ and comparator-dependent, with no robust
sign. The $-0.22$ appears only against one baseline-flow shadow (an outlier, $\sim2\times$
every other matched-amplitude drive), but is $+0.03$ against the loop-isolating replay and
$\approx0$ against a persistence-matched telegraph and the continuum grid. Matching the
herd-side first-moment amplitude leaves its higher-moment (burst) structure unmatched. The
result is asymmetric: one robust observable with a genuine self-reinforcing reflexive
component under momentum and an ambiguous one under OFI. The dry-up's magnitude, separately,
follows a smooth amplitude$\times$persistence continuum.

Two features are scoped precisely. The effect remains a smooth crossover, not a
discontinuous Dark Corner, confirmed at fine $\varphi$ resolution over three $\kappa$ rows.
The RMS-mispricing growth toward high $\varphi$ remains a placement/dilution
artefact--largest with imitation off, suppressed by real imitation--which we decline to
call a Dark Corner. The contribution is therefore a disciplined, robust, mechanistically
resolved result: a real emergent liquidity crossover, distinguished by an explicit null, a
rule/horizon-robustness battery, and an amplitude-matched mechanism decomposition, both from
the placement artefact and from exogenous-shock flash-crash framings
\citep{gao2022flashcrash}.

\section*{Code and data availability}
The code and data reproducing all results reported here are available in a public
repository, \coderepo.

\bibliographystyle{unsrtnat}
\bibliography{references}

\end{document}